\documentclass[12pt]{article}
\usepackage{hyperref}
\usepackage{a4,amsmath,srcltx,bbm,dutchcal}%
\usepackage{amssymb,amsfonts,graphicx}%
\usepackage{float}%
\usepackage[notlof]{tocbibind}%
\usepackage{cite}%
\usepackage[titletoc,title]{appendix}%
\usepackage{color,soul}%
\usepackage{authblk}%

\author[1]{Yimin Jiang}
\affil[1]{\textit{\small{Central South University, 410083 Changsha, China}}}
\author[2]{Itai Einav}
\affil[2]{\textit{\small{School of Civil Engineering, The University of Sydney, 2006 Sydney, Australia.}}}
\author[3]{Mario Liu}
\affil[3]{\textit{\small{Theoretische Physik, Universit{\"a}t T\"{u}bingen, 72076 T\"{u}bingen, Germany}}}

\begin{document}
\bibliographystyle{unsrt}%
\title{A Thermodynamic Treatment\\ of  Partially Saturated Soils \\ Revealing the Structure of Effective Stress}
\maketitle

\begin{abstract}
A rigorous thermodynamic treatment of partially saturated soils is developed using a minimal number of assumptions. The derivation is carried out in a way that does not require to explicitly track the complex shapes of interfaces between the solid, fluid and gas domains. Instead, suction is the property being recovered explicitly through the minimisation of energy around an ideal `suctionless limit', while considering the different compressibilities of the three domains. In interpreting experimental data the derivation ensures the thermodynamic equilibrium between the chemical potentials of the soil and measurement cells, while carefully distinguishing intrinsic from measured pressures and suctions. A most general expression for the effective stress of partially saturated soils is then derived that is strictly linked to the soil-water retention curve (SWRC). The structure of the effective stress broadly depends on the three thermodynamic densities characterising the solid, fluid and gas domains. Special cases of SWRC are explored, which reveals conditions for which the structure of the effective stress may agree with previously proposed empirical relationships. 
\end{abstract}

\newpage
\tableofcontents
\newpage
\noindent

%

\section{Introduction}
Partially saturated soils are mixtures of solid, liquid and gas domains. Unlike situations in fully saturated and fully dry soils, in partially saturated soils the interfaces between the domains take non-trivial shapes and suction develops. Many attempts have been made to thermodynamically consider such soil mixtures. 

The seminal work of Gibbs \cite{Gibbs1878} on the chemical thermodynamics of molecularly dispersed substances introduced the unequivocal definition of chemical potential of a given substance as the variable conjugate to the density of that substance,  the partial derivative of the Helmholtz free energy density with respect to the concerned density, at constant temperature $T$, ie.  $\mu\equiv\partial f/\partial\rho|_T$. Day \cite{Day42} adopted those ideas to soil mixtures for interpreting moisture measurements and further (tacitly) postulated that this potential could be decomposed into separate decoupled terms. Similar ideas were then advanced in soil sciences and physics \cite{Day42,Edlefsen43,Babcock57,Sposito81}. For example, Sposito \cite{Sposito81} assumed that the chemical potential could be decoupled into a sum of so-called matrix, pneumatic, and envelope-pressure potentials, while Babcock and Overstreet \cite{Babcock57} further split the matrix potential into so-called capillary-pressure and adsorption potentials. However, the tacit decoupling of all these terms may be questioned, and therefore a more general and rigorous thermodynamic treatment is required. 

Whereas soil scientists tend to interpret soil moisture measurements using potentials, soil mechanicians usually opt to use pressure-like quantities such as measured suction, and air and water pressures, as those are pertinent for mechanical stress calculations. However, the `measured suction' -- the difference between the air and water pressures in measurement cells -- should not be confused with the difference of those pressures in the soil mixture itself. As highlighted in \cite{Baker2009} some constitutive models have incorrectly identified measured suction with the capillary component of the matrix potential of the soil. It should also be noted that the matrix potential is not a true thermodynamic potential since it depends on the decoupling assumption. A rigorous thermodynamic treatment is therefore needed that distinguishes between intrinsic and measured pressures, and between intrinsic and measured suctions, while maintaining the equilibrium between the chemical potentials in the soil mixture and measurement cells.  

In describing the effective mechanical behaviour of partially saturated soils many thermodynamic works have advanced the description of internal constitutive properties in the soil mixture ({\it e.g.}, see {\cite{Muraleetharan99,Hutter99,Gray01,Schrefler02,Laloui03,deBoer2005,Gray10,Coussy10,Zhao10,Nikooee12,Buscarnera12} to name a few). For example, some papers attempt to track the complex shapes and distribution of interfaces, which are not easily accessible experimentally. Surface tensions are typically homogenised to recover the measured suction, with an inevitable loss in accuracy. The current paper avoids this complex strategy by inverting the logic. This is done by introducing the measured suction explicitly into the thermodynamic description, such that the role of interfaces on the effective material behaviour could be captured  implicitly though more accurately.

An effective stress of partially saturated media is meant to replace the role of total stress in constitutive models of dry materials with no air pressure. For example, according to Bishop \cite{Bishop59} the effective stress of partially saturated soils $\sigma^{eff}_{ij}$ takes the following form:
\begin{equation}
\sigma^{eff}_{ij}\equiv\sigma_{ij}-P_T\delta_{ij},\,\,\,\,\,  P_T=u_A-\chi(u_A-u_W),
\label{eq:Bishop}
\end{equation}
where $\delta_{ij}$ is the Kronecker delta tensor, $\sigma_{ij}$ the total stress tensor, $u_A$ and $u_W$ are the measured air and water pressures, $s=(u_A-u_W)$ the measured suction, and $P_T$ the effective pressure (in this paper, to be interpreted as a thermodynamic pressure) that acts as a weighted average of $u_A$ and $u_W$ with a weighting factor $\chi$ (the so-called Bishop's parameter). In fully saturated soils ($S_r=1$), where the measured suction vanishes ($s=0$), $u_A=u_W$ and therefore $P_T=u_W$, in agreement with Terzaghi's \cite{Terzaghi43} effective stress principle for such soils. This is true for any $\chi$, although some take $\chi=1$ for $S_r=1$. However, in partially saturated soils ($S_r\neq 1$), there is no agreement on the form of $\chi$. 

Many forms have been proposed for $\chi$, mostly based on empirical arguments. It is frequently assumed that $\chi$ depends solely on the degree of saturation $S_r$, while arguments have also been motivated for $\chi=S_r$\cite{Hassanizadeh80,Bear84,Houlsby97}. The sole dependence of $\chi$ on $S_r$ has been questioned empirically \cite{JenBurland62}, with some suggesting $\chi$ should instead be solely dependent on the measured suction $s$ \cite{Aitchison85,Khalili98} (with $s$ itself being generally a function of both $S_r$ and solid density \cite{Assouline06,Tarantino09,Romero11,Zhou12,AlHaj16}). Expressions of $\chi$ solely dependent on $s$ or $S_r$ have been suggested to be different for shear strength and volumetric compression ({\it e.g.}, \cite{Nuth08}), which involve different changes to solid density during the experiments. Such expressions have also been modified for double porosity soils with morphologically different distributions of macro and micro pores \cite{Borja09JMPS,Bagherieh09}. To date, thermodynamics have not been demonstrated to explain such diverse empirical observations on $\chi$ for various soils and along different loading conditions. This is one of the goals of the current paper. 

Specifically, the validity of Bishop's effective stress principle in Eq.(\ref{eq:Bishop}) will be demonstrated, with $\chi$ generally dependent on the three thermodynamic densities characterising the solid, liquid and gas domains. The expression for $\chi$ is shown to be strictly connected to the soil-water retention curve (SWRC), which supports the conclusion of \cite{Alonso10} based on empirical and micro-structural arguments. Special cases of SWRC are explored, which reveal when $\chi$ may agree with the previously proposed relationships above. However, the formulation is not restricted to these special cases of SWRC. As such, it may be used to capture the influence of phenomena such as SWRC hystereses during wetting and drying cycles~\cite{Kool1987,Masin2010} on the effective stress.

The current treatment deviates from the considerations of classical thermodynamics such as given by Truesdell and others\cite{TruesdellNoll1965,Truesdell1965}. Here, the derivation follows the {\it hydrodynamic procedure} by Landau, as presented in the books on Newtonian and superfluid liquids~\cite{LandauL5,LandauL6,Khalatnikov}. It was generalized to liquid crystals by de Gennes\cite{deGennes}, and more recently to granular materials by Jiang and Liu\cite{granR2}. Note that both the first and second laws of thermodynamics, as well as Truesdell's physical constraints of material objectivity and symmetry, are fully contained in the hydrodynamic procedure. More crucially, this more comprehensive hydrodynamic approach enables one to derive the thermodynamic pressure, which we will employ to obtain the structure of the effective stress.

\section{The Basic Physics} \subsection{The State Variables\label{1}}
We consider a mixture of solid grains (S), liquid (such as water, W) and gas (such as air, A) that, though finely dispersed, consists of  single-component domains {\it sufficiently macroscopic for thermodynamics to hold in each of them}. 
For given temperature, the mixture is characterised by the volume $V$ and masses, $M_S, M_W$ and $M_A$; equivalently, it can be characterised by the thermodynamic partial densities 
\begin{equation}
\varrho_S\equiv M_S/V,\,\, \varrho_W\equiv M_W/V,\,\, \varrho_A\equiv M_A/V.
\end{equation}
The Helmholtz free energy (per unit volume) is a function of these densities, 
\[f=f(\varrho_S,\varrho_W,\varrho_A,\cdots).\] 
By defining the thermodynamic total density, and water and air concentrations
\begin{equation}
\varrho\equiv\varrho_S+\varrho_W+\varrho_A,\,\, c_W\equiv\varrho_W/\varrho,\,\,c_A\equiv\varrho_A/\varrho,
\end{equation}
we may equally take $f=f(\varrho,c_W,c_A,\cdots)$. 
As each of the three systems occupies a subvolume, $V_S,V_W,V_A$, with $V=V_S+V_W+V_A$, the averaged true or {\it intrinsic densities} and the single-component free energies are
\begin{equation}
\hat\varrho_\beta\equiv M_\beta/V_\beta,\quad \hat f_\beta=\hat f_\beta(\hat\varrho_\beta),\quad\beta=S,W,A.
\end{equation}
Finally,  the volume ratios $\phi_\beta$, porosity $n$ and degree of saturation $S_r$ are defined as
\begin{align}\label{4}
\phi_\beta\equiv\frac{V_\beta}V=\frac{\varrho_\beta}{\hat\varrho_\beta},\qquad\quad \sum_\beta\phi_\beta=1,\qquad\\\nonumber  n\equiv\frac{V-V_S}{V}=\phi_W+\phi_A,\,\,\,\, S_r\equiv\frac{V_W}{V-V_S}=\frac{\phi_W}{n}.
\end{align}
Next, recall that the thermodynamic pressure ($P_T$, where here the subscript "$T$" is used to highlight "thermodynamic") of liquid or gas is given by the change of free energy $F$  in the rest frame with respect to that of the volume $V$: 
\begin{equation}
P_T\equiv-\frac{\partial F}{\partial V}=-\left.{\frac{\partial (f V/M)}{\partial (V/M)}} \right\vert_{M}=-\frac{\partial (f/\varrho)}{\partial (1/\varrho)}=-f\frac{\partial(1/\rho)}{\partial(1/\rho)}+\frac{\rho^2}{\rho}\frac{\partial f}{\partial\rho}=-f+\rho\mu,
\end{equation} 
where the overall chemical potential is denoted as $\mu\equiv\partial f/\partial\rho$. Also note that we assume constant temperature, $T=$ const, and thus neglect in the following the dependence of $f$ on $T$ and thus the possible dependence of the effective stress on temperature. However, such dependencies could be quite easily included in the future. Therefore, given the Helmholtz free energies, $f$ and $\hat f_\beta$, the associated chemical potentials and pressures are
\begin{align}\label{6}
\mu_\beta\equiv\partial  f/\partial\varrho_\beta,\quad
\mu\equiv\partial  f/\partial\varrho,\quad
P_T\equiv\varrho\mu-f.
\\
\hat\mu_\beta\equiv\partial\hat  f_\beta/\partial\hat\varrho_\beta,\qquad
\hat P_\beta\equiv\hat\varrho_\beta\hat \mu_\beta-\hat f_\beta.\qquad
\end{align}
The hatted symbols denote {\it intrinsic} quantities, which characterise the single-component domains, the non-hatted are {\it thermodynamic} quantities characterising the mixture. 

Denoting the total stress as $\sigma_{ij}$, we have $\sigma_{ij}=P_T\delta_{ij}=\hat P_A\delta_{ij}$  in air in equilibrium, and  $\sigma_{ij}=P_T\delta_{ij}=\hat P_W\delta_{ij}$ in a Newtonian fluid. The sign convention for the stress is taken consistently with respect to the momentum balance, $\partial_t g_i+\nabla_j (\sigma_{ij}+g_iv_j)=0$, with $g_i$ being the momentum and $v_i$ the velocity, as adopted in Appendix~\ref{sec:AppA}.
More generally, especially for partially saturated soil, the total stress is given by 
\begin{equation}\label{StressDecom}
\sigma_{ij}=\sigma_{ij}^{e}+P_T\delta_{ij},
\end{equation}
where the elastic stress, from the soil skeleton,  is given as $\sigma_{ij}^{e}=-\partial f/\partial \varepsilon^{e}_{ij}$, with $\varepsilon^{e}_{ij}$ denoting the elastic strain. In Appendix~\ref{sec:AppA}, we give a brief thermodynamic validation of this expression, showing that it is the only one  compatible with energy and momentum conservation, and the second law of thermodynamics. 

Because with an appropriate expression for $f$, the stress $\sigma_{ij}^{e}$ is well capable of accounting for static deformation and elasto-plastic motion of dry granular media\cite{granRexp}, we identify $\sigma_{ij}^{e}$ with  $\sigma_{ij}^{eff}$ of Eq.(\ref{eq:Bishop}). 
%
We note that the expression $\sigma_{ij}^{e}=-\partial f/\partial \varepsilon^{e}_{ij}$ holds for arbitrarily large total strain $\varepsilon_{ij}$ -- such as given in an approach to the critical state. However, the definition of the elastic strain in Eq. (\ref{3y-GSH}) should be redefined for large deformations by those interested to model non-equilibrium problems. 
We also note that all three domains possess an average intrinsic pressure, $\hat P_S,\hat P_W,\hat P_A$, with $\hat P_S$ denoting the response of the solid domain being compressed isotropically by air and water. It is non-zero even if the solid domain is fragmented into non-contacting grains. Therefore,  we take $\hat P_S$ and $\sigma_{ij}^{eff}$ to be independent from  each other.

Two further remarks. First, $M_\beta$ and $V$ can be changed at will, but $V_\beta$ adjusts for given $M_\beta$ and $V$, as a result of the force equilibrium between the single-component systems. The force equilibrium between the three components is comparatively quickly established and represents the basic interaction of the mixture. Therefore, we take $\varrho_\beta$ as independent thermodynamic variables, and $\hat\varrho_\beta$ as the dependent ones. This is why the free energy $f$ is a function of $\varrho_\beta$ only. 

Second, $M_W,M_A$ and $M_S$ are assumed conserved. This is usually a good approximation because at atmospheric pressure the air dissolved in water at 20$^\circ$ C is $\lesssim$ 2\%, and the vapour concentration in air at 30$^\circ$ is $\approx$ 3\%. But this assumption may be dramatically wrong for specific circumstances, say close to the boiling point at given pressure. Then $M_W$ will be divided into a liquid and a gaseous part, such that the chemical potential of the water is equal to that of the vapour.
This additional complication will be addressed in a future work.

Starting from the infinitesimal Helmholtz free energy ${\rm d}f$ at constant  temperature, 
\begin{equation}
{\rm d}f=\mu_S{\rm d}\varrho_S +\mu_A{\rm d}\varrho_A +\mu_W{\rm d}\varrho_W-\sigma_{ij}^{eff}{\rm d}\varepsilon^{e}_{ij},
\end{equation}
with 
$c_\beta\equiv\varrho_\beta/\varrho$, $\varrho\equiv\sum\varrho_\beta$,
$\sum c_\beta=1$, ${\rm d}\varrho_A={\rm d}(\varrho c_A)=\varrho{\rm d}c_A +c_A{\rm d}\varrho$, and  $\mu_S{\rm d}\varrho_S +\mu_A{\rm d}\varrho_A=\mu_S{\rm d}(\varrho_S+\varrho_A) +(\mu_A-\mu_S){\rm d}\varrho_A$, we also have 
\begin{align}\label{10}
{\rm d}f=&\mu{\rm d}\varrho +\chi_A{\rm d}c_A +\chi_W{\rm d}c_W-\sigma_{ij}^{eff}{\rm d}\varepsilon^{e}_{ij},
\\\nonumber
&\chi_A\equiv\varrho(\mu_A-\mu_S),\quad \chi_W\equiv\varrho(\mu_W-\mu_S),
\\\nonumber
&\mu\equiv\sum\mu_\beta c_\beta=\sum\mu_\beta\varrho_\beta/\varrho,
\\\nonumber
&P_T\equiv\varrho\mu-f=\textstyle{\sum}\varrho_\beta\mu_\beta-f.
\end{align}
Note $\varrho\mu={\sum}\varrho_\beta\mu_\beta$, which is in fact a result of a more general formula. With $\varrho\frac{\partial F}{\partial\varrho}=\varrho\sum\frac{\partial F}{\partial\varrho_\alpha}\frac{\partial\varrho_\alpha}{\partial\varrho}=\varrho\sum c_\alpha\frac{\partial F}{\partial\varrho_\alpha}=\sum \varrho_\alpha\frac{\partial F}{\partial\varrho_\alpha}$ for any $F=F(\varrho_\alpha)$ (not only the free energy), we have 
\begin{equation}\label{11}
\varrho\frac{\partial }{\partial\varrho}=\sum_\alpha \varrho_\alpha\frac{\partial }{\partial\varrho_\alpha}.
\end{equation}


\subsection{The Intrinsic Suction}
If, instead of the mixture, we have the case of three large single domains, see Fig.(\ref{fig1}), force equilibrium  implies  equal pressures,
\begin{equation}
P_0\equiv \hat P_\beta,  \quad \beta=S,W,A.
\end{equation}
This equilibrated pressure $P_0$ is here termed as the `common pressure'. This should remain valid if  the solid parts / grains are sufficiently large. 
\begin{figure}[t]
\centering
\includegraphics[scale=.7]{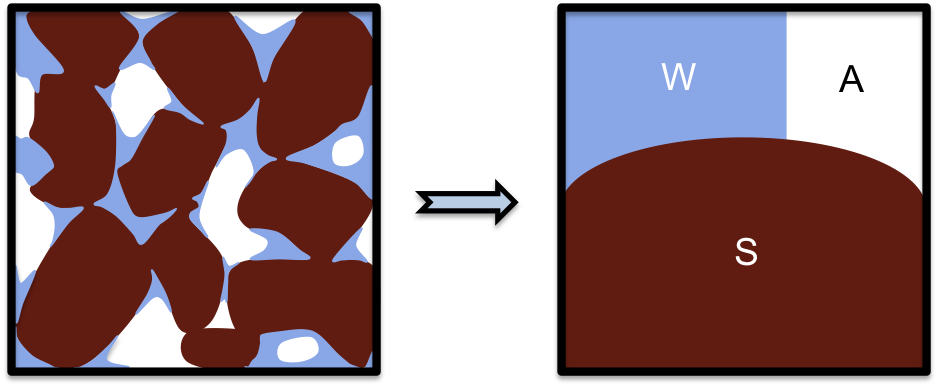}
\caption{A mixture, characterised by its three thermodynamic densities, $\varrho_S,\varrho_W$ and $\varrho_A$, can have different distributions of its single-component domains. The left figure has suction, the right, with three large single-component domains, does not. Yet if the left system did not have suction, both would have the same relationships between the thermodynamic and intrinsic quantities, such as density or pressure. 
To account for suction, we expand the pressures in the small changes of the intrinsic densities: $\hat\varrho_S, \hat\varrho_W$ and $\hat\varrho_A$, around the suctionless state.}\label{fig1}
\end{figure}
For smaller grains, one observes a deviation from the common pressure $P_0$, 
\begin{equation}\label{13}
\Delta\hat P_\beta=\hat P_\beta-P_0,
\end{equation}
which grows as the grain size decreases. Hereby, the air pressure increases and the water pressure reduces. This is due to the smaller surface energy for solid-water interfaces than solid-air interfaces, which implies reduced interfaces with air. 
Therefore, air bubbles, semi-bubbles, or regions form between the grains covered with water. The smaller the grains, the larger the surface curvature between water and air, producing a surface tension that compresses the air. 
(The capillary force that draws the water up the narrow gaps between grains, against the gravitation,  is also a result of the differing surface energies.)
The phenomenon of reduced water pressure is frequently called {\it suction}. 
The {\it intrinsic suction} is the difference between the intrinsic air and water pressures:
\begin{equation}\label{intrinsic suction}
\hat{s}=\hat P_A-\hat P_W=\Delta\hat P_A-\Delta\hat P_W. 
\end{equation}

\subsection{The Measured Suction\label{2}}
Stiff porous walls can be manufactured such that only water or only air penetrate through them, such that the cell on the other side contains only water or air, as illustrated in Fig.~(\ref{fig2}). Where thermodynamic equilibrium is being guaranteed, it can be shown that this scheme represents the conditions of various experimental techniques. 
For a given temperature, the cell fluids are characterised by their respective density $\varrho^{\rm cell}_\beta$,  free energy $f^{cell}_\beta\equiv\hat f_\beta(\varrho^{\rm cell}_\beta)$, chemical potential $\mu_\beta^{cell}$ and pressure $u_\beta$, where 
\begin{equation}
\mu_\beta^{cell}=\hat\mu_\beta(\varrho^{\rm cell}_\beta),\,\, u_\beta=\hat P_\beta(\varrho^{\rm cell}_\beta),\quad  \beta=A,W.
\end{equation}
Having measured $u_A$ and $u_W$, we take 
\begin{equation}\label{16}
{s\equiv u_A-u_W}\,
\end{equation}
to be the {\it measured suction}, the quantity that we shall relate to the thermodynamic pressure $P_T$ in this work. 
This is possible because on one hand, $P_T$ is a function of $\varrho_A,\varrho_W$ and $\varrho_S$, and on the other,
equilibrium with respect to mass transfer between the mixture and cells implies 
\begin{align}\label{15}
{\mu_\beta(\varrho_A,\varrho_W,\varrho_S)=\mu_\beta^{\rm cell}(\varrho^{\rm cell}_\beta)},
\end{align}
which determines $\varrho_\beta^{\rm cell}$, and with them also the two cell pressures $u_\beta=\hat P_\beta(\varrho^{\rm cell}_\beta)$.

Finally, it should be highlighted that the intrinsic and measured suctions are not identical, as highlighted in Fig.(\ref{fig2}).

\begin{figure}[H]
\centering
\includegraphics[scale=.8]{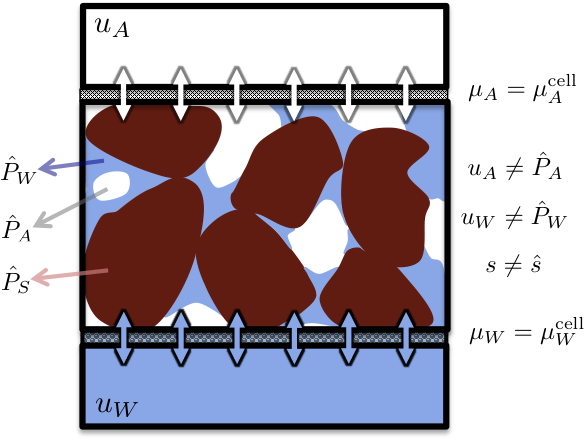}
\caption{Thermodynamic equilibrium along walls requires the chemical potentials in the soil ($\mu_A$ and $\mu_W$) to match those in the measurement cells ($\mu_A^{cell}$ and $\mu_W^{cell}$). It does not require equality between the intrinsic (hatted quantities) and measured (non-hatted) pressures. Subsequently, we must distinguish between intrinsic ($\hat s=\hat P_A-\hat P_W$) and measured ($s=u_A-u_w$) suctions.}\label{fig2}
\end{figure}


\subsection{The Suctionless Limit and the Terzaghi Principle\label{Terz}}

In the present treatment that is thermodynamic, we shall not derive $\Delta\hat P_\beta$ for a given granular geometry and the surface energies between solid, water and air. 
Instead, we shall relate both $P_T$ and the measured suction $s=u_A-u_W$ to  $\Delta\hat P_\beta$. This allows us to infer the expressions for $P_T$ from $s$. Then, given Eq.(\ref{StressDecom}), it is easy to obtain $\sigma_{ij}^{eff}$. 
In this approach, the `suctionless limit', epitomised by three large single domains, is useful. For this case,  we expect: $P_T=P_0$, as this is the common pressure of all domains. (The notion of partial pressures, $P_T=\sum\hat P_\beta$, as in air, does not hold here, because air consists of several gases that are microscopically dispersed and  do not interact.)
Also, in the suctionless limit there is no difference between the cell pressure and the respective single component intrinsic pressure, and therefore $\hat P_\beta=u_\beta$ holds. It follows that in the suctionless limit
\begin{equation}\label{17}
P_T= P_0=\hat P_\beta=u_\beta,\quad \hat\varrho_\beta=\varrho_\beta^{cell}.
\end{equation}
This has two ramifications. First, considering Eq.(\ref{StressDecom}) with $P_T=u_W=u_A$, 
\begin{equation}\label{18}
\sigma^{eff}_{ij}=\sigma_{ij}-P_T\delta_{ij}=\sigma_{ij}-u_W\delta_{ij}=\sigma_{ij}-u_A\delta_{ij}.
\end{equation}
This is the Terzaghi principle -- for any saturation, not only  $S_r=1$. On the other hand, if there is only water and indeed $S_r=1$, even small grains will not develop suction, as there is no granular contact with air that can be reduced. We therefore conclude 
\begin{equation}\label{19}
\Delta\hat P_\beta\to0\quad\text{for}\quad S_r\to1. 
\end{equation}
These limits need to be reproduced by the calculations in the following Sec.\ref{sec3}. 

Using the same logic, we have $\Delta\hat P_\beta\to0$ also for $S_r\to0$, implying no measured suction $s=u_A-u_W\to0$, and not the observed  $s\to\infty$. We do not understand this discrepancy yet, but believe it is related to cohesion ({\it i.e.}, intermolecular attractive force through the water film covering and compressing the grains). Cohesion is not included in the current treatment but will be added in the near future.

The second ramification of Eqs.(\ref{17}) is $\mu_\beta(\varrho_\beta)
=\mu_\beta^{\rm cell}(\varrho^{\rm cell}_\beta) =\hat\mu_\beta(\hat\varrho_\beta)$. (The first equal sign is always valid, see Eq.(\ref{15}); the second holds because the densities are equal.) 
With suction, only $\mu_\beta=\mu_\beta^{\rm cell}$ remains, while generally $\mu_\beta^{\rm cell}\not=\hat\mu_\beta$. 
This subtle point is highlighted since when suction develops some take $\varrho^{\rm cell}_\beta=\hat\varrho_\beta$, $u_\beta=\hat P_\beta$, $s=\Delta\hat P_A-\Delta\hat P_W$, which implies erroneously that $\mu_\beta^{\rm cell}=\hat\mu_\beta$ still holds. As mentioned above, the property $\Delta\hat P_A-\Delta\hat P_W$ is the intrinsic suction $\hat s$, and not the measured suction $s$, as highlighted in Fig.~(\ref{fig2}).

\subsection{The Strategy of the Derivation}
Given the  intrinsic Helmholtz free energies and densities $\hat f_\beta,\hat \rho_\beta$, it is easy to calculate the associated chemical potentials $\hat\mu_\beta=\partial\hat f_\beta/\partial\hat \rho_\beta$ and pressures $\hat P_\beta=\hat\mu_\beta\hat\rho_\beta-\hat f_\beta$. 
Similarly, given the Helmholtz free energy and partial densities, $f,\rho_\beta$, the expressions for $\mu, \mu_\beta$ and $P_T=\mu\rho-f$ follow. One can then go on to obtain the cell pressures, $u_A,u_W$, and the Bishop's stress factor $\chi=(P_T-u_A)/(u_W-u_A)$. These are all the steps one needs to follow in order to define the effective stress of partially saturated porous media.    

However, the actual problem is not as straightforward because during experiments only the three masses $M_\beta$ and total volume $V$ are known, but not the partial volumes $V_\beta$. This implies that one knows the partial densities $\rho_\beta$, but not the intrinsic densities $\hat\rho_\beta$. If there was no suction, the three  $\hat\rho_\beta$ are fixed by two force equilibrium conditions, $\hat P_A(\hat\rho_A)=\hat P_W(\hat\rho_W)=\hat P_S(\hat\rho_S)$, in addition to the constraint $V=\sum V_\beta=\sum M_\beta/\hat\rho_\beta$. Where suction exists (and $\Delta\hat P_\beta\not=0$), the last constraint still holds, but the two force equilibrium  conditions are modified. Together, they again yield $\hat\rho_\beta$. (Again, only two of $\Delta\hat P_\beta$ are independent, as the third is given by $ \sum V_\beta=V$.)

The information on intrinsic suction, $\hat s=\Delta\hat P_A-\Delta\hat P_W$,  as a function of the three partial densities $\rho_\beta$,  depends sensitively on the curvatures of the interfaces, and their size distribution, which is obviously not easy to obtain. But this information is also encoded in the SWRC, ie. the measured suction $s\equiv u_A-u_W$. 
To utilise this information, we may simply postulate an $\hat s(\rho_\beta)$ such that the calculated SWRC is similar to the observed/measured one. 

In addition to the postulated intrinsic suction $\hat s$, we need a second condition to fix the two independent $\Delta\hat P_\beta$. 
One may assume that the intrinsic solid pressure remains unchanged, $\Delta\hat P_S=0$, though this  would be an arbitrary, \textit{ad hoc} step.  A much better possibility is to minimize the free energy $f(\rho_\beta,\hat\rho_\beta)$ with respect to the dependent variables $\hat\rho_\beta$. This approach is physically convincing since the minimization of the free energy (equivalent to maximisation of the entropy) yields what is realised in nature with overwhelming probability.

If the minimization is done without any constraints, the two Euler-Lagrange equations are given by the conditions of force equilibrium, $\hat P_A=\hat P_W=\hat P_S$. Minimizing with the constraint that $\Delta\hat P_A-\Delta\hat P_W$ is a given function $\hat s$, the one resultant Euler-Lagrange equation is the second condition that we are looking for. 
   
To execute the described calculation, we need the explicit expressions for $f$ and $\hat f_\beta$. The intrinsic energy for air, water, and granular bulk material, $\hat f_A, \hat f_W$ and $\hat f_S$  are in principle well known, though they may be nonlinear under extreme pressure. And we take $f$ to be a sum of  $\hat f_\beta$, weighted by their volume fraction.  To obtain simple formulas, rid of any familiar (and hence unnecessary) complications, we expand $\hat f_\beta$ to linear order around a state that we refer to as 1 -- such that any nonlinearities are in 1 rather than the expanded expressions.   

This simplification alone is sufficient to yield fully analytic solutions, though the final expressions are complex and very long. Therefore, we employ a second simplification, an expansion around the suctionless limit explained in Sec:\ref{sec3}.  Hereby, because the first-order terms turn out to be exceedingly small, we need to go to second-order terms, which are simple, transparent and meaningful. These two solutions have been compared numerically and are practically identical, and thus we only present the latter, more transparent solution.
\section{The Free Energy\label{sec3}}

The overall free energy $Vf$ is quite generally given by the sum of the intrinsic free energies, $Vf=\sum V_\beta\hat f_\beta$, or
\begin{equation}\label{22}
f=\sum \phi_\beta\hat f_\beta(\hat\varrho_\beta)=\sum({\varrho_\beta}/{\hat\varrho_\beta})\hat f_\beta\,.
\end{equation}
Surface contributions to $f$, which are the reason leading to suction, are not {\it explicitly} considered. But of course, they cannot be neglected. Therefore, here suction will be the quantity accounted for explicitly in Sec.~\ref{InclSuct} by minimising $f$ with constraints, in a way that {\it implicitly} considers the surface effects on the effective behaviour in an experimentally accessible manner.  

More specifically, the interfacial energy, being two-dimensional in nature, is always dominated by the three dimensional bulk energies -- this holds in spite of its considerable extension. The reason it cannot be neglected is because it gives rise to an extra force (or surface tension), large enough to be well noticeable in the force balance. If one wants to calculate the surface tension, there is no easy way to avoid the interfacial energy, and with it the usually complex geometry of the soil sample under consideration. On the other hand, if one only wants to take the measured suction, and correlate it to the force imbalance of the sample (both being the result of the same interfacial energy and soil geometry), it is sufficient to postulate the force imbalance by a constraint and calculate the measured suction, as is done in the following. This remains valid irrespective of the physical origin of the force imbalance, whatever the sample geometry or the value of the interfacial energy is. Yet it should be pointed out that the results may depend, in general, though apparently not within our approximations, on the constraints adopted in the following Sec.~\ref{InclSuct}.

We also neglect the temperature dependence of $f$ in this paper, as it is not central to accounting for the phenomenon of suction. If temperature dependence would be needed, it is sufficient to include it in  $\hat f_\beta=\hat f_\beta(T,\hat\varrho_\beta)$.

With $f$  given and  $\hat f_\beta$ arbitrary, the pressure 
$P_T\equiv\varrho({\partial f}/{\partial\varrho})-f=\sum\varrho_\beta({\partial f}/{\partial\varrho_\beta})-f$ and chemical potentials $\mu_\beta\equiv{\partial f}/{\partial\varrho_\beta}$  are (see details in App.\ref{sec:AppB})
\begin{align}\label{P_T}
P_T
&=\sum_\beta  \hat P_\beta\frac{\varrho\,\varrho_\beta}{\hat\varrho_\beta^2}
\left.\frac{\partial\hat\varrho_\beta}{\partial\varrho}\right|_{c_\beta}=\sum_\beta \hat P_\beta\left[1-\varrho\frac{\partial}{\partial\varrho}\right]\phi_\beta
\\\nonumber 
&=\sum_\beta  \hat P_\beta\frac{\varrho_\beta}{\hat\varrho_\beta^2}
\sum_\alpha\varrho_\alpha\frac{\partial\hat\varrho_\beta}{\partial\varrho_\alpha}
=\sum_{\beta}\hat P_\beta\left[1-\sum_\alpha\varrho_\alpha\frac{\partial}{\partial\varrho_\alpha}\right]\phi_{\beta},
\\\label{mu}
\mu_\beta
&=\hat \mu_\beta-\frac{\hat P_\beta}{\hat\varrho_\beta}+\sum_\alpha \varrho_\alpha\frac{\hat P_\alpha}{\hat\varrho_\alpha^2}\frac{\partial\hat\varrho_\alpha}{\partial\varrho_\beta}
=\hat \mu_\beta-\sum_\alpha \hat P_\alpha
\frac{\partial\phi_\alpha}{\partial\varrho_\beta}.
\end{align}
The importance of the free energy in Eq.(\ref{22}) should therefore be highlighted, since it enabled us to relate $P_T$ to $\hat P_\beta$, and $\mu_\beta$ to $\hat\mu_\beta$, and thus the property of the mixture to that of the well-understood single domains.

Note that for the three second equal signs one of these relations were used: 
\begin{equation}
\varrho\frac{\partial\phi_\beta}{\partial\varrho}=\phi_\beta\left[1-\frac{\varrho}{\hat\varrho_\beta}\frac{\partial\hat\varrho_\beta}{\partial \varrho}\right],\qquad
\hat\varrho_\beta\frac{\partial\phi_\beta}{\partial\varrho_\alpha}=\left[\delta_{\alpha\beta}-\phi_\beta\frac{\partial\hat\varrho_\beta}{\partial \varrho_\alpha}\right],
\end{equation} 
where $\delta_{\alpha\beta}$ is the unity matrix.

\subsection{The Variation of the Free Energy Including Suction}\label{InclSuct}
The free energy density $f$ has, initially, six variables: $\varrho_\beta, \hat\varrho_\beta$. The latter three are rendered  functions of the former by minimising  $\int f{\rm d}^3r$ for given $\varrho_S,\varrho_W,\varrho_A$, and for $\sum_\beta\phi_\beta=1$, {\it cf.} Eq.(\ref{4}). Taking $L_1$ as the constant Lagrange parameter, the variational calculation reads
\begin{align}\label{23}
\delta\int& (f+L_1\sum_\beta\phi_\beta){\rm d}^3r=\sum_\beta\delta\int(\hat f_\beta+  L_1)\phi_\beta\,{\rm d}^3r
\\\nonumber&=\sum_\beta\int\left[\frac{\partial\hat f_\beta}{\partial\hat\varrho_\beta}{\phi_\beta}+(\hat f_\beta+  L_1)\frac{\partial\phi_\beta}{\partial\hat\varrho_\beta}\right]\,{\delta\hat\varrho_\beta}\,{\rm d}^3r
\\\nonumber
&=\sum_\beta\int\left[\hat\mu_\beta-(\hat f_\beta+ L_1)\frac1{\hat\varrho_\beta}\right]\,\phi_\beta\,{\delta\hat\varrho_\beta}\,{\rm d}^3r
\\\nonumber&=\sum_\beta\int\left[\hat P_\beta(\hat\varrho_\beta)-  L_1\right]\frac{\phi_\beta}{\hat\varrho_\beta}\,{\delta\hat\varrho_\beta}\,{\rm d}^3r=0.
\end{align}
Since each of $\delta\hat\varrho_\beta$ varies independently, the minimisation condition is: 
\begin{equation}\label{24}
\hat P_\beta(\hat\varrho_\beta)=L_1=P_0.
\end{equation}
This case corresponds to the suctionless case given by Eq.(\ref{17}). We therefore need to introduce another constraint, which will enable pressure differences $\Delta\hat P_\beta\equiv\hat P_\beta-P_0$ and suction to develop. 

Pressure differences and suction are of course the result of the contribution of surface free energy. This is most simply accounted for by an additional constraint for the above variational calculation. An obvious choice is to add the intrinsic air-water pressure difference $s_{AW}=\hat P_A-\hat P_W$, with $s_{AW}$ being a function of $\varrho_\beta$ (such that Eq.(\ref{19}) holds), but not of  $\hat\varrho_\beta$. 
Taking $L_2$ as a second Lagrange parameter leads to $\delta\int (f-L_1 \sum_\beta \phi_\beta-L_2[\hat P_A-\hat P_W]){\rm d}^3r=0$, or 
\begin{align}
\label{s_AW}
\hat P_S=\hat P_W+L_2 K_W/\phi_W&
=\hat P_A-L_2 K_A/\phi_A=L_1=P_0+\Delta P_0,
\\\nonumber
\text{implying}\quad s_{AW}&=L_2(K_A/\phi_A+K_W/\phi_W),
\end{align} 
where the bulk modulus $K_\beta$ is introduced as
\begin{equation}\label{K}
K_\beta\equiv\hat\varrho_\beta\times\partial\hat p_\beta/\partial\hat\varrho_\beta.
\end{equation}
In the next section, Sec.\ref{sec2.1}, Eqs.(\ref{33}), $K_\beta$ will be approximated as density-independent. Since $K_W$ is much larger than $K_A$, see Eq.(\ref{eq:Ks}), the water pressure increases much more than the air pressure. Note that due to the second constraint, $L_1$ in Eq.(\ref{s_AW}) is not equal to $L_1$ in Eq.(\ref{24}). Taking  $P_0$ as the suctionless value, we need  $\Delta P_0$ to account for the difference. 

Two alternative constraints could be explored
\begin{align}
\label{25}
\hat P_W&=L_1-s_W,\quad\text{or}
\\\label{25b}
\hat P_A&=L_1+s_A,  
\end{align} 
with the first approximating Eq.(\ref{s_AW}), and the second approximating its (unrealistic) opposite. As the first constraint completely fixes $\hat P_W$, it implies that there is no variation with respect to $\delta\hat\varrho_W$, so the minimisation simply yields $\hat P_A=\hat P_S=L_1$. The second constraint, analogously, fixes $\hat P_A$. Hence
\begin{align}\label{s_W}
\hat P_A=\hat P_S=\hat P_W+s_W&=P_0+\Delta P_0,\quad\text{or}
\\\label{s_A}
\hat P_W=\hat P_S=\hat P_A-s_A&=P_0+\Delta P_0.
\end{align}
In the following we will show that the first and second alternative constraints are indeed realistic (giving either $\hat s=s_{AW}$ or $\hat s=s_W$), while the third alternative constraint being unrealistic.

\subsection{The Intrinsic Free Energies\label{sec2.1}} 

Next, we specify the intrinsic free energies, $\hat f_\beta$, for the bulk solid, water and air, respectively. These are in principle well-known. However, for the simplicity and transparency of the results, we shall expand all three intrinsic pressures around a reference value $P_1$ (say of 1 atm),
\begin{equation}
\Delta\hat P_\beta\equiv\hat P_\beta - P_0,\quad P_0=P_1+\delta P.
\end{equation}
The total deviation from $P_1$ is separated into two parts: $\delta P$ as the universal change that occurs by changing $\varrho_\beta$ in a  suctionless system ({\it e.g.}, changing $M_\beta$ for $V=$ const), and $\Delta\hat P_\beta$ as the additional change from turning on suction at given $\varrho_\beta$ (for $M_\beta,V=$ const). 
This notation applies to other quantities as well, {\it e.g.},
\begin{equation}
\Delta\hat\varrho_\beta\equiv\hat\varrho_\beta-\hat\varrho_\beta^0,\quad \hat\varrho_\beta^0\equiv\hat\varrho_\beta^1+\delta\hat\varrho_\beta^0. 
\end{equation}
Expanding the intrinsic pressure $\hat P_\beta$ in its density $\hat\varrho_\beta$ to first order in $\Delta\hat\varrho_\beta$, or $P_0$ to first order in $\delta\hat\varrho_\beta^0$, implies a linear relationship between both. We note that the suction-induced changes, $\Delta\hat\varrho_\beta$, should be rather limited. First, the suction is not strong enough to appreciably change $\hat\varrho_W$ or $\hat\varrho_S$. Although air is compressible, $\Delta\hat\varrho_A$ is also small, because $\sum\Delta\phi_\beta=0$. Therefore,  a calculation to linear order should  suffice. 

If we consider geological pressures, $\delta P$ may be considerable. But we could always shift $P_1$ from 1 atm, such that   $\delta P\equiv P_1-P_0$ is small enough for a linear expansion in $\delta\hat\varrho_W$ and $\delta\hat\varrho_S$. Even though $\delta\hat\varrho_A$ will be much larger, $\delta P\sim\delta\hat\varrho_A$ holds as long as the ideal gas law does. Therefore, we take (see Eq.(\ref{K}))
\begin{align}\label{33}
\delta P=K_\beta\frac{\delta\hat\varrho_\beta}{\hat\varrho_\beta}=K\frac{\delta\varrho}{\varrho},\quad \Delta\hat P_\beta=K_\beta\frac{\Delta\hat\varrho_\beta}{\hat\varrho_\beta},\qquad\quad
\\\nonumber
\text{where }\quad K_\beta\equiv\hat\varrho_\beta\partial P/\partial \hat\varrho_\beta=-V_\beta\partial P/\partial V_\beta\quad 
\end{align}
We note that Eqs.(\ref{33}) is equivalent to specifying the free energy as $\Delta\hat f_\beta=K_\beta\Delta\hat \varrho^2_\beta/2\hat \varrho^2_\beta$, with $\Delta\hat\mu_\beta=K_\beta{\Delta\hat\varrho_\beta}/ {\hat\varrho_\beta}^2$, and  $\Delta\hat P_\beta=\hat \varrho_\beta\Delta\hat\mu_\beta-\Delta\hat f_\beta=K_\beta{\Delta\hat\varrho_\beta}/ {\hat\varrho_\beta}$ to linear order in ${\Delta\hat\varrho_\beta}$.
The values for the bulk modulus vary widely, 
\begin{equation}\label{eq:Ks}
K_A=10^5\text{Pa},\,\, K_W=2\cdot10^9\text{Pa},\,\, K_{\rm glass}=3\cdot10^{10}\text{Pa},\,\, K_{\rm steel}=2\cdot10^{11}\text{Pa}.
\end{equation}
Similarly, the associated density vary widely, $2\cdot10^4\times{\delta\hat\varrho_W}/{\hat\varrho_W}={\delta\hat\varrho_A}/{\hat\varrho_A}$, 
with ${\delta\hat\varrho_S}/{\hat\varrho_S}$ smaller by at least another order of magnitude. Yet since the densities (in {kg/m}$^3$ at 1 atm) are also quite different:   
\begin{equation}\label{35}
\hat\varrho_A=1,\,\, \hat\varrho_W=10^3,\,\, \hat\varrho_{\rm glass}=2.5\cdot10^3, \,\, \hat\varrho_{\rm steel}=\cdot10^4.
\end{equation}
The values for their combined effect $K_\beta/\varrho_\beta$ are better aligned,
\begin{equation*}
\frac{K_A}{\varrho_A}=\frac1{20}\frac{K_W}{\varrho_W}=\frac1{200}\frac{K_{glass}}{\varrho_{glass}}.
\end{equation*}	

\subsection{The Common Pressure} 
The last section enables one to obtain the common pressure $P_0$ for any mixture of a given volume $V$ and three masses $M_\beta$. We start from three single-component systems of the same masses, and presume that their densities $\hat\varrho_\beta$ at $P_1=$1 atm are known. Their volumes are then $V_\beta=M_\beta/\hat\varrho_\beta$, and the sum is  $V_1=\sum V_\beta$.  Given the difference between the two volumes  $\delta V=V-V_1$, we obtain the difference in the pressure $\delta P=P_0-P_1$ by $\delta P=-(K/V)\delta V$, or
\begin{equation}\label{P_0}
P_0=P_1-(K/V)(V-V_1).
\end{equation}
To obtain the total bulk modulus $K$, we 
vary $\varrho$ and $V$ at fix $M_\beta$, employing Eqs.(\ref{33}),
\begin{align}\nonumber
\delta V=\sum\delta V_\beta&=\sum M_\beta\delta(1/\hat\varrho_\beta)
=-\sum  V_\beta(\delta\hat\varrho_\beta/\hat\varrho_\beta)=-\delta P\sum(V_\beta/K_\beta),
\\
K&=\varrho\left.\frac{\partial P}{\partial\varrho}\right|_{c_\beta}=-V\frac{\partial P}{\partial V}
=V\left/\sum\frac{V_\beta}{K_\beta}\right. \label{KK}
\end{align}
For $V_A/V\gg K_A/K$, and since  air is the only compressible component, we have to first order in $\varepsilon\equiv V_WK_A/V_AK_W$ and $\bar\varepsilon\equiv V_SK_A/V_AK_S$, 
\begin{equation}\label{41*}
\frac KV\approx \frac{K_A}{V_A}\left(1-\frac{V_W}{V_A}\frac{K_A}{K_W}-\frac{V_S}{V_A}\frac{K_A}{K_S}\right)
\equiv\frac{K_A}{V_A}(1-\varepsilon-\bar\varepsilon).
\end{equation}
For a fully saturated system, $V_A=0$, we have 
\begin{equation}
\frac KV=\left[\frac{V_W}{K_W}+\frac{V_S}{K_S}\right]^{-1}.
\end{equation}

Finally, noting Eqs.(\ref{4},\ref{33}), we use 
\[0=\sum\Delta\phi_\alpha=-\sum\phi_\alpha\Delta\hat\varrho_\alpha/\hat\varrho_\alpha=\sum V_\alpha\Delta\hat P_\alpha/VK_\alpha\] 
to find $\Delta\hat P_A=0$, to zeroth order in $\varepsilon,\bar\varepsilon$.  
Inserting this into Eqs.(\ref{s_AW},\ref{s_W},\ref{s_A}), we find $\Delta P_0=0$ for the first two cases, and  $\Delta P_0=-s_A$ for the third case, or 
\begin{align}\label{58} 
\Delta\hat P_W&=-s_{AW},\quad \Delta\hat P_A=\Delta\hat P_S=0,
\\\label{56} 
\Delta\hat P_W&=-s_W,\quad \Delta\hat P_A=\Delta\hat P_S=0,
\\\label{57} 
\Delta\hat P_W&=\Delta\hat P_S=-s_A,\quad \Delta\hat P_A=0.
\end{align}
With the first two identical, and the third unrealistic, we shall from here on only consider the second possibility. Therefore, the intrinsic suction is given as
\begin{equation}\label{intS} 
\hat s = s_W=-\Delta\hat P_W.
\end{equation}

A possible extension may be carried out by avoiding to take the zeroth order approximation in $\varepsilon,\bar\varepsilon$, in which case the above result will depend on all the three domain compressibilities.

\subsection{The Intrinsic densities for the Suctionless Case\label{sec2.3}}
With   $\delta\hat\varrho_\beta=\hat\varrho_\beta\delta P/K_\beta$ and $\phi_\beta={\varrho c_\beta}/{\hat\varrho_\beta}$, we have 
\begin{align}\label{41}
\left.\frac{\partial\hat\varrho_\beta}{\partial\varrho}\right|_{c_\beta}&=\frac{\hat\varrho_\beta}{\varrho}\frac {K}{K_\beta}\approx\frac{\hat\varrho_\beta}{\varrho}\frac{V}{V_A}\frac{K_A}{K_\beta}=\frac{\hat\varrho_\beta}{\varrho}\frac{\hat\varrho_A}{\varrho_A}\frac{K_A}{K_\beta},
\\\nonumber
\left(1-\varrho\frac{\partial}{\partial\varrho}\right)\phi_\beta&=\frac{\varrho\varrho_\beta}{\hat\varrho_\beta^2}\frac{\partial\hat\varrho_\beta}{\partial\varrho}=\frac{V_\beta}V\frac{K}{K_\beta}\approx\frac{V_\beta}{V_A}\frac{K_A}{K_\beta}.
\end{align}
To calculate $\delta\hat\varrho_\alpha/\delta\varrho_\gamma$ at given $\varrho_\beta$, $\beta\not=\gamma$, we vary $M_\gamma$ leaving $M_\beta$ and $V$ constant. With
$0=\delta V=\sum\delta V_\alpha=\sum\delta(M_\alpha/\hat\varrho_\alpha)=\delta M_\gamma/\hat\varrho_\gamma-\sum V_\alpha(\delta\hat\varrho_\alpha/\hat\varrho_\alpha)$, 
{we have}
$\delta M_\gamma/\hat\varrho_\gamma=V\delta\varrho_\gamma/\hat\varrho_\gamma=\sum V_\alpha (\delta\hat\varrho_\alpha /\hat\varrho_\alpha) =(V/K)\delta P
=(V /K)K_\alpha\delta\hat\varrho_\alpha /\hat\varrho_\alpha$, or
\begin{align}\label{42}
\frac{\partial\hat\varrho_\alpha}{\partial\varrho_\gamma}&=\frac{\delta\hat\varrho_\alpha}{\delta\varrho_\gamma}=\frac{\hat\varrho_\alpha}{\hat\varrho_\gamma}\frac{K}{K_\alpha}\approx\frac{\hat\varrho_\alpha}{\hat\varrho_\gamma}\frac{V}{V_A}\frac{K_A}{K_\alpha}
=\frac{\hat\varrho_\alpha}{\hat\varrho_\gamma}\frac{\hat\varrho_A}{\varrho_A}\frac{K_A}{K_\alpha},
\\\label{43}
\hat\varrho_\beta\frac{\partial\phi_\alpha}{\partial\varrho_\beta}&=\delta_{\alpha\beta}-\frac{\hat\varrho_\beta\varrho_\alpha}{\hat\varrho_\alpha^2}\frac{\partial\hat\varrho_\alpha}{\partial\varrho_\beta}={\delta_{\alpha\beta}}-\frac{V_\alpha}{K_\alpha}
\frac{K}V\approx
\delta_{\alpha\beta}-\frac{V_\alpha}{K_\alpha}\frac{K_A}{V_A}.
\end{align}
To obtain ${\hat\varrho_\beta=\hat\varrho_\beta(\varrho_\gamma)}$, we start from a reference pressure (say, as mentioned, 1 atm) of given volume $V$ and $M_\alpha$. The thermodynamic densities $\varrho_\alpha^1=M_\alpha/V$ and the intrinsic densities $\hat\varrho_\beta^1$ are presumed known, as they are simply the densities of the three single-component systems at the reference pressure. To linear order from it, we have, from Eq.(\ref{42}),
\begin{equation}\label{45*}
\frac{\hat\varrho_\beta-\hat\varrho_\beta^1}{\hat\varrho_\beta^1}
=\frac{\hat\varrho_A^1}{\varrho_A^1}\frac{K_A}{K_\beta}\sum_\gamma\frac{\varrho_\gamma-\varrho_\gamma^1}{\hat\varrho_\gamma^1}.
\end{equation}
where the summation comes from the integration constant and the fact that $\gamma$ needs to be varied. Clearly, only $\hat\varrho_A$ changes appreciably, doing so equally with all three densities.

\subsection{An Approximation Scheme for the Suction Case}
The above is a complete theory, and the equations may be solved analytically, with the help of a computer. What one does is to first relate $\Delta\hat P_W=-\hat s$ to ${\Delta\hat\varrho_\beta}$ using Eqs.(\ref{33}), then insert the result into Eqs.(\ref{P_T},\ref{mu}), to obtain $P_T$ and $\mu_\beta$. Given $\mu_\beta$, we also have the knowledge of $u_\beta$, see Sec.\ref{2}. This gives us a relation between $P_T$ and $s=u_A-u_W$. 
Unfortunately, the end expressions extend over pages, and are  too complicated to be illuminating. 
Therefore, for transparency we employ below an approximation scheme that simplifies the solution. 

First, given the linear relation $\Delta\hat P_\beta\sim{\Delta\hat\varrho_\beta}$, Eq.(\ref{33}), we shall also confine the relations between the pressures and chemical potentials to linear order  in ${\Delta\hat\varrho_\beta}$, 
\begin{align}\label{p-mu}
\hat\varrho_\beta^0\Delta\hat\mu_\beta=\Delta\hat P_\beta,
\quad
\hat\varrho_\beta^0\Delta\mu^{cell}_\beta&=\hat\varrho_\beta^0\Delta\mu_\beta=\Delta u_\beta.
\end{align}
With Eq.(\ref{15}), the measured suction is therefore
\begin{equation}\label{eq:msuc}
s\equiv u_A-u_W=\Delta u_A-\Delta u_W=\hat\varrho_A^0\Delta\mu_A-\hat\varrho_W^0\Delta\mu_W.
\end{equation}
Next, we consider the thermodynamic pressure and chemical potentials. For the suctionless case, $\hat P_\beta=P_0$, and thus the expressions for $P_T$ and $\mu_\beta$ in  Eqs.(\ref{P_T},\ref{mu}) reduce to 
\begin{align*}
P_T&=\sum_\beta \hat P_\beta\left[\phi_\beta-\varrho\frac{\partial\phi_\beta}{\partial\varrho}\right]=P_0\sum_\beta (1-0)=P_0,
\\ \mu_\beta&=\hat \mu_\beta^0-\sum_\alpha \hat P_\alpha
\frac{\partial\phi_\alpha}{\partial\varrho_\beta}=\hat \mu_\beta^0- \hat P_0\frac{\partial}{\partial\varrho_\beta}\sum_\alpha\phi_\alpha=\hat \mu_\beta^0,
\end{align*}
as expected in Eqs.(\ref{15},\ref{17}). Subtracting these from Eqs.(\ref{P_T},\ref{mu}), we find 
\begin{align}\label{D P_T1}
\Delta P_T&
=\sum_\beta \Delta \hat P_\beta\left[1-\varrho\frac{\partial}{\partial\varrho}\right]\phi_\beta
=\sum_\beta \Delta  \hat P_\beta\frac{\varrho\,\varrho_\beta}{\hat\varrho_\beta^2}
\frac{\partial\hat\varrho_\beta}{\partial\varrho}=
\sum_{\alpha,\beta} \Delta  \hat P_\beta\frac{\varrho_\alpha\varrho_\beta}{\hat\varrho_\beta^2}
\frac{\partial\hat\varrho_\beta}{\partial\varrho_\alpha},
\\\label{D P_T2}
\Delta \mu_\alpha&
=\Delta \hat \mu_\alpha-\sum_\beta \Delta \hat P_\beta
\frac{\partial\phi_\beta}{\partial\varrho_\alpha}=
\sum_\beta\Delta \hat P_\beta \frac{\varrho_\beta}{\hat\varrho_\beta^2}\frac{\partial\hat\varrho_\beta}{\partial\varrho_\alpha},\qquad\quad \text{implying}
\\\label{FinP_T}
\Delta P_T&=\sum\phi_\alpha{\hat\varrho_\alpha}\Delta \mu_\alpha=\sum\phi_\alpha\Delta u_\alpha,\quad\text{or}\quad P_T=\sum\phi_\alpha u_\alpha,
\end{align}
where $u_S\equiv P_0+{\hat\varrho_S}\Delta \mu_S$ is used for shorthand, as it is of course as yet not measurable. 

 In computing $\Delta P_T$ and $\Delta \mu_\alpha$ to first order in ${\Delta\hat\varrho_\beta}$, we employ Eqs.(\ref{D P_T1},\ref{D P_T2}), because $\Delta\hat P_\beta\sim{\Delta\hat\varrho_\beta}$, while it should be sufficient to include only the zeroth order term of the Jacobian matrix  ${\partial\hat\varrho_\beta}/{\partial\varrho_\alpha}$, {\it i.e.}, taking its suctionless value as given in  Sec.\ref{sec2.3}. Unfortunately, this does not work because  $\Delta P_T$ and $\Delta \mu_\beta$ are exceedingly small, implying that second order terms $\sim(\Delta\hat P_W)^2\sim \hat s^2$ dominate. This is why, in a second step,  we include in ${\partial\hat\varrho_\beta}/{\partial\varrho_\alpha}$ the terms linear in $\sim\Delta\hat P_\beta$. However, we do not go beyond Eq.(\ref{33}) or (\ref{p-mu}), as the first order terms here are large, with the second order ones being tiny corrections.  

Although this derivation is slightly subtle and formal, the derived approximation is transparent and consistent, and we always have the computer-generated exact solution as a check.

\subsection{First Order Solution  for the Suction Case\label{sec3.4}}
Noting Eqs.(\ref{41},\ref{43}), we calculate the first order corrections to the thermodynamic pressure $\Delta P_T=P_T-P_0$ and chemical potentials $\Delta\mu_\beta$ of Eqs.(\ref{D P_T1},\ref{D P_T2}),
\begin{align}\label{44}
\Delta P_T
=&\sum_\beta \Delta\hat P_\beta\left[1-\varrho\frac{\partial}{\partial\varrho}\right]\phi_\beta
=\sum_\beta \Delta\hat P_\beta\frac{V_\beta}V\frac{K}{K_\beta},
\\\nonumber
\hat\varrho_\beta^0\Delta\mu_\beta&=\hat\varrho_\beta^0\Delta\hat \mu_\beta-\sum_\alpha\Delta \hat P_\alpha\,
\hat\varrho_\beta\frac{\partial\phi_\alpha}{\partial\varrho_\beta} 
\\\nonumber 
&=\hat\varrho_\beta^0\Delta\hat \mu_\beta-\sum_\alpha\Delta \hat P_\alpha\left[{\delta_{\alpha\beta}}-\frac{V_\alpha}{K_\alpha}\frac{K}V\right]
=\sum_\alpha\Delta \hat P_\alpha\frac{V_\alpha}{K_\alpha}\frac{K}V.
\end{align}
Since $\Delta u_\beta=\hat\varrho_\beta^0\Delta\mu_\beta$ does not depend on $\beta$, see  Eq.(\ref{p-mu}), this implies, first of all,  
\begin{equation}\label{45}
s=\hat\varrho_A^0\Delta\mu_A-\hat\varrho_W^0\Delta\mu_W=0,
\end{equation}
an obvious contradiction to observation. 
In $P_T$,  only  $\Delta\hat P_A$ contributes in zeroth order of $\varepsilon$. Including contributions of first order in $\varepsilon$ and $\bar\varepsilon$, see Eq.(\ref{41*}), we have:
\begin{equation}\label{47} 
P_T=P_0+\Delta\hat P_A+(\Delta\hat P_W-\Delta\hat P_A)\varepsilon+(\Delta\hat P_S-\Delta\hat P_A)\bar\varepsilon.
\end{equation}
Inserting Eq.(\ref{56}) yields
\begin{equation}
P_T=P_0 + {\cal O}(\varepsilon^2).
\end{equation}
Clearly, the first-order terms are so small, that they are dominated by the second order terms, and we need to go beyond the zeroth order of ${\partial\hat\varrho_\beta}/{\partial\varrho_\alpha}$ in Eqs.(\ref{D P_T1},\ref{D P_T2}).

\subsection{Second Order Solution  for the Suction Case}
Including the first-order terms in the Jacobian Matrix (while ignoring the zeroth order ones), 
\[\frac{\partial\hat\varrho_\beta}{\partial\varrho_\alpha}=\frac{\partial(\hat\varrho_\beta^0+\Delta\hat\varrho_\beta)}{\partial\varrho_\alpha}=\frac{\partial\hat\varrho_\beta^0}{\partial\varrho_\alpha}+\frac{\hat\varrho_\beta^0}{K_\beta}\frac{\partial \Delta\hat P_\beta}{\partial\varrho_\alpha}\approx \frac{\hat\varrho_\beta^0}{K_\beta}\frac{\partial \Delta\hat P_\beta}{\partial\varrho_\alpha},\] 
and inserting it  into Eqs.(\ref{D P_T1},\ref{D P_T2}) yields  $u_\alpha=\hat\varrho_\beta\Delta\mu_\alpha$:
\begin{align}\label{55}
\Delta u_\alpha=\sum_\beta\left[\frac{\hat\varrho_\alpha\phi_\beta}{2K_\beta}\frac{\partial (\Delta\hat P_\beta)^2}{\partial\varrho_\alpha}\right].
\end{align}
Continuing with Eq.(\ref{intS}) yields
\begin{align}\label{59} 
{\Delta u_\alpha=\frac{\phi_W}{2K_W}\hat\varrho_\alpha \psi_\alpha},\quad {\psi\equiv \hat s^2,\quad \psi_\alpha\equiv\frac{\partial\psi}{\partial \varrho_\alpha}}.
\end{align}
We proceed with this second order solution, which provides an accurate framework to assess the structure of the effective stress in the next section. Using Eq.~(\ref{eq:msuc}) we further highlight the difference (yet also the strict connection through $\psi$) between the intrinsic ($\hat s$) and measured ($s$) suctions
\begin{align}\label{eq:twosuctions}
{\hat s=\sqrt{\psi}},\quad {s=\frac{\phi_W}{2K_W}(\hat\varrho_A \psi_A-\hat\varrho_W \psi_W)}.
\end{align}


\section{The Effective Stress }
\subsection{General structure}
The general structure of the effective stress is known using Eq.(\ref{StressDecom}), which depends on the thermodynamic pressure given by Eq.(\ref{FinP_T})
\[P_T=P_0+\sum_\alpha \phi_\alpha\Delta u_\alpha=u_A+\sum\phi_\alpha(u_\alpha-u_A).\] 
By rearranging the last equation we can always (formerly) express $P_T$ in a Bishop form with $\chi=\phi_W+\phi_S (u_A-u_S)/(u_A-u_W)$. Thus, considering Eq.~(\ref{59}), we find that for any general $\psi\equiv\psi(\varrho_A,\varrho_W,\varrho_S)$ 
\begin{align}\label{eq:GENpt}
&P_T=u_A-\chi(u_A-u_W),
\\\label{eq:GENchi}
&\chi\equiv\chi(\varrho_A,\varrho_W,\varrho_S)=\phi_W+\phi_S\left[\frac{\hat\varrho_A \psi_A-\hat\varrho_S \psi_S}{\hat\varrho_A \psi_A-\hat\varrho_W \psi_W}\right],
\\\label{eq:GENs}
&s\equiv s(\varrho_A,\varrho_W,\varrho_S)=\frac{\phi_W}{2K_W}(\hat\varrho_A \psi_A-\hat\varrho_W \psi_W).
\end{align}
Note that in this general case the measured suction $s$ depends on all the thermodynamic densities, including the air density $\varrho_A$. Considering the links between Eqs. (\ref{59},\ref{eq:twosuctions}) suggests the SWRC depends on $u_A$, even for fixed suction. Note that in soil mechanics the SWRC curves are typically recovered experimentally by adopting the {\it axis translation} concept that requires one to neglect any dependence of $s$ on $u_A$. However, some authors have questioned the validity of this simplification \cite{Or02,Baker2009}. 

More generally, the above equations present an intimate relationship between $\chi$ and $s$ ({\it i.e.}, the SWRC), through their mutual dependence on $\psi$, which supports the observation of \cite{Alonso10} based on empirical and micro-structural arguments. This relationship will be highlighted explicitly using the examples in the next section.

\subsection{Special cases}
Many forms of soil-water retention curves have been proposed in the literature (see for example \cite{Brooks64,Genuchten80,Gallipoli03,Assouline06,Tarantino09,Romero11,Zhou12,AlHaj16}, just to name a few), each employing different primary variables. In the following we explore how the choice of the primary variables affects the reduced structures of Bishop parameter in Eq.~(\ref{eq:GENchi}) and measured suction in Eq.~(\ref{eq:GENs}).
\\[12pt](i)\indent Suction independent on air density.\\[6pt]
Since in this case $\hat s\equiv\hat s(\varrho_W,\varrho_S)$, and in general $\psi=\hat s^2$, it follows that $\psi_A=0$ and thus

\begin{align}\label{casei}
\chi\equiv\chi(\varrho_W,\varrho_S)=\phi_W+\phi_S\left[\frac{\hat\varrho_S \psi_S}{\hat\varrho_W \psi_W}\right], \quad s\equiv s(\varrho_W,\varrho_S)=-\frac{\varrho_W \psi_W}{2K_W}.
\end{align}
Since the measured suction $s$ is also independent on $\varrho_A$, it follows that in this case the experimental {\it axis translation} technique should work. 
\\[12pt](ii)\indent Suction dependent on water density.\\[6pt]
Given that $\hat s\equiv\hat s(\varrho_W)$, both $\psi_A=0$ and $\psi_S=0$. Therefore, Eq.(\ref{casei}) reduces to
\begin{align}\label{60}
\chi\equiv\chi(\varrho_W)=\phi_W=nS_r,\quad s\equiv s(\varrho_W)=-\frac{\varrho_W \psi_W}{2K_W}.
\end{align}
Therefore, in this case, when $S_r=1$ we find $\chi=n$, but in that limit there should be no measured suction $s=0$, thus $u_A=u_W$, so the structure of the effective stress would still agree with Terzaghi's principle for fully saturated soils.
\\[12pt](iii)\indent Suction dependent on degree of saturation and porosity.\\[6pt]
In this case $\hat s\equiv\hat s(S_r,n)$. Thus, by denoting $\psi_{sr}={\partial\psi}/{\partial S_r}$ and $\psi_n={\partial\psi}/{\partial n}$, we find
\begin{align}\nonumber
{\Delta u_\alpha}&=\frac{\phi_W}{2K_W}\left[\hat\varrho_\alpha \frac{\partial S_r}{\partial\varrho_\alpha} \psi_{sr}-\delta_{\alpha S}\psi_n\right],
\\\nonumber
\frac{\partial S_r}{\partial\varrho_\alpha}&=-(S_r)^2\cdot\frac{\partial(\phi_A/\phi_W)}{\partial\varrho_\alpha}
=\frac{(S_r)^2}{\phi_W}\,\,\frac{1}{\hat\varrho_\alpha}\left[\frac{\phi_A}{\phi_W}{\delta_{\alpha W}}+1-{\delta_{\alpha A}}\right].
\end{align}
Note that in order to calculate ${\partial S_r}/{\partial\varrho_\alpha}$, we employed Eq.(\ref{42}) of the suctionless limit, because the quantities $\psi,\psi_{sr},\psi_n$ are already of second order. Also, we only include terms  to zeroth order in $\varepsilon,\bar\varepsilon$, and hence took only $\hat\varrho_A$ to depend on $\varrho_\beta$ (the other intrinsic densities are approximately constant). It follows that $\Delta u_A=0$, $\Delta u_W=\frac{S_r}{2K_W}\psi_{sr}$, and $\Delta u_S=\Delta u_W S_r-\frac{\phi_W}{2K_W}\psi_n$, from which we find
\begin{align}\label{eq:srn}
\chi\equiv\chi(S_r,n)=S_r-n(1-n)\frac{\psi_n}{\psi_{sr}},\quad s\equiv s(S_r,n)=-\frac{S_r \psi_{sr}}{2K_W}.
\end{align}
Therefore, the sign of $\psi_n/\psi_{sr}$ determines whether $\chi$ would be larger or smaller than $S_r$. 
\\[12pt](iv)\indent Suction dependent on degree of saturation.\\[6pt]
In this fourth case of $\hat s\equiv\hat s(S_r)$, $\psi_n=0$, such that using the previous case
\begin{align}
\chi\equiv\chi(S_r)=S_r,\quad s\equiv s(S_r)=-\frac{S_r \psi_{sr}}{2K_W}.
\end{align}
The result of $\chi=S_r$ (as suggested by many authors through other reasonings \cite{Hassanizadeh80,Bear84,Houlsby97}) is therefore thermodynamically consistent with SWRC that are strictly dependent on $S_r$, but independent on $n$. However, the SWRC does most generally depend on both $S_r$ and $n$, as clearly demonstrated experimentally ({\it e.g.}, \cite{Gallipoli03,Tarantino09}). 
%

\subsection{Examples}
Next, we examine how the shape of the soil-water retention curve (SWRC) influences the Bishop's effective stress factor $\chi$. 
Towards this aim a special function is explored for $\psi$ below. However, the formulation is not restricted to this choice only. Other choices could be easily explored, and may include for example the influence of other SWRC shape factors and phenomena such as SWRC hystereses during wetting and drying cycles. Without loss of generality, and only for demonstration purposes, consider therefore the following form for $\psi$:

\begin{equation}
\psi=2A K_W \phi_S^{\beta}\left[\frac{1-\alpha+\alpha S_r-S_r^\alpha}{\alpha (1-\alpha)S_r^\alpha}\right],
\end{equation}
with $\alpha$, $\beta$ and $A$ being model parameters. Since $\phi_S=1-n$, this example belongs to case (iii) above. We can therefore use Eq.~(\ref{eq:srn}) to find
\begin{align}
s=(A \phi_S^{\beta}) \left[\frac{1-S_r}{S_r^{\alpha}}\right],\quad
\chi=S_r-\beta{\phi_W}\left[\frac{1-\alpha+\alpha S_r-S_r^\alpha}{\alpha (1-\alpha)(1-S_r)}\right].
\end{align}
The terms including $\phi_S$ and $S_r$ in $s$ are multiplicatives. Therefore, in this model the dependence of the SWRC on $S_r$ is controlled by the factor $\frac{1-S_r}{S_r^{\alpha}}$ through the shape parameter $\alpha$ (see Fig.(\ref{fig3}a)), while $A \phi_S^\beta$ captures the apparent `air entry value' with parameter $\beta$ adjusting the effect of solid density on that value (see Fig.(\ref{fig3}b)). For positive $\beta$ the apparent air entry value increases with increasing solid density, which is more realistic than the opposite effect of negative $\beta$. Also, when $\beta=0$ the air entry value is simply given by $A$, and Bishop parameter becomes $\chi=S_r$, as expected from cases (iii) and (iv) above. 

Finally, note that this model has a logarithmic limit when $\alpha\to 1$,
\begin{gather}
\psi=2A K_W \phi_S^\beta[\ln S_r+S_r^{-1}-1],
\\
s=(A \phi_S^{\beta}) [S_r^{-1}-1],\quad
\chi=S_r-\beta{\phi_W}\left[1+\frac{\ln S_r}{S_r^{-1}-1}\right].
\end{gather}
In this $\alpha\to 1$ limit, when $\beta=0$ the soil-water retention curve $s=A [S_r^{-1}-1]$ reduces to the one previously proposed by \cite{Buscarnera12}.

Fig.(\ref{fig3}) demonstrates the ability of the model to capture a wide range of SWRC by changing the two parameters $\alpha$ and $\beta$, and the porosity $n$. Figs.~(\ref{fig4}) and (\ref{fig5}), and (\ref{fig6}) illustrate the effects of $\beta$, $n$ and $\alpha$, respectively, on the Bishop parameter $\chi$. While $\beta$ and $n$ change the shape of the $\chi$-$S_r$ curves (see Figs.~(\ref{fig4}a and (\ref{fig5})a, respectively), they do not affect the principal shape in the $\chi$-$s$ space (Fig.~(\ref{fig4}b and (\ref{fig5}b, respectively), here always showing an asymptote of about 0.55, as previously proposed empirically by \cite{Khalili98}. On the other hand, $\alpha$ does not have a strong effect on the shape of the $\chi$-$S_r$ curves (Fig.(\ref{fig6}a)), but does affect strongly the shape in the $\chi$-$s$ space (Fig.(\ref{fig6}b). In this case, the lines approach a variety of possible asymptotes, including the asymptote of 0.55 as in \cite{Khalili98} and of 1 as in \cite{Aitchison85}.

\begin{figure}[H]
\centering
\includegraphics[scale=.52]{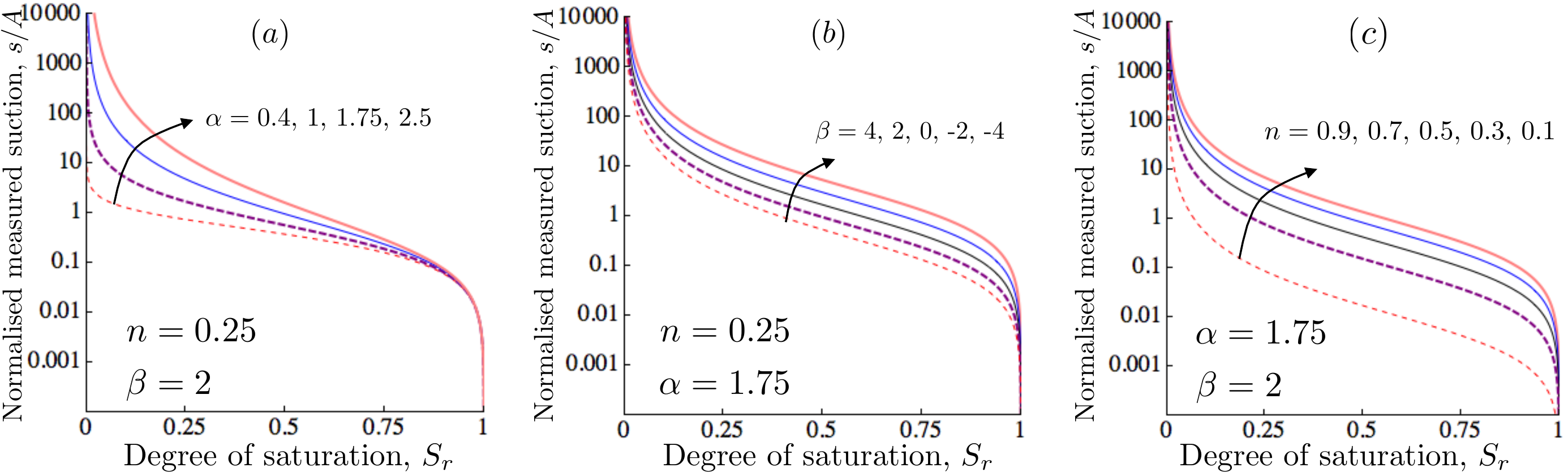}
\caption{A family of soil-water retention curves (SWRC). (a) The case of $\beta=2$ and $n=0.25$ ($\phi_S=0.75$), for various $\alpha$'s, which shows how $\alpha$ controls the SWRC's dependence on degree of saturation $S_r$; (b) The case of $\alpha=1.75$ and $n=0.25$ ($\phi_S=0.75$), for various $\beta$'s; (c) The case of $\alpha=1.75$ and $\beta=2$, for various fixed porosity values of $n$.}\label{fig3}
\end{figure}

\begin{figure}[H]
\centering
\includegraphics[scale=.65]{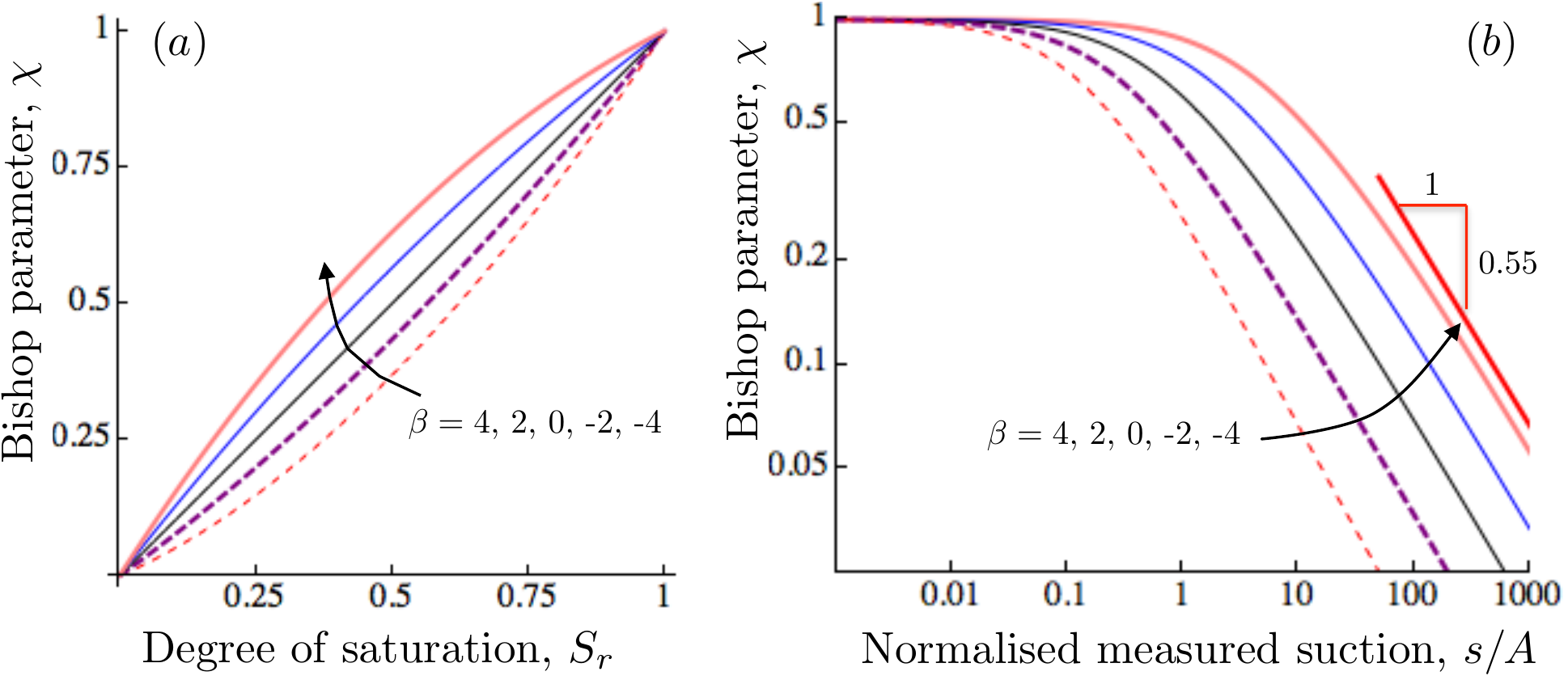}
\caption{The effect of $\beta$ on Bishop parameter $\chi$, for fixed $\alpha=1.75$ and $n=0.25$. (a) Bishop parameter plotted in terms of the degree of saturation; (b) Bishop parameter plotted in terms of measured suction, agreeing with the relation of \cite{Khalili98} that has an empirical asymptote of 0.55.}\label{fig4}
\end{figure}

\begin{figure}[H]
\centering
\includegraphics[scale=.65]{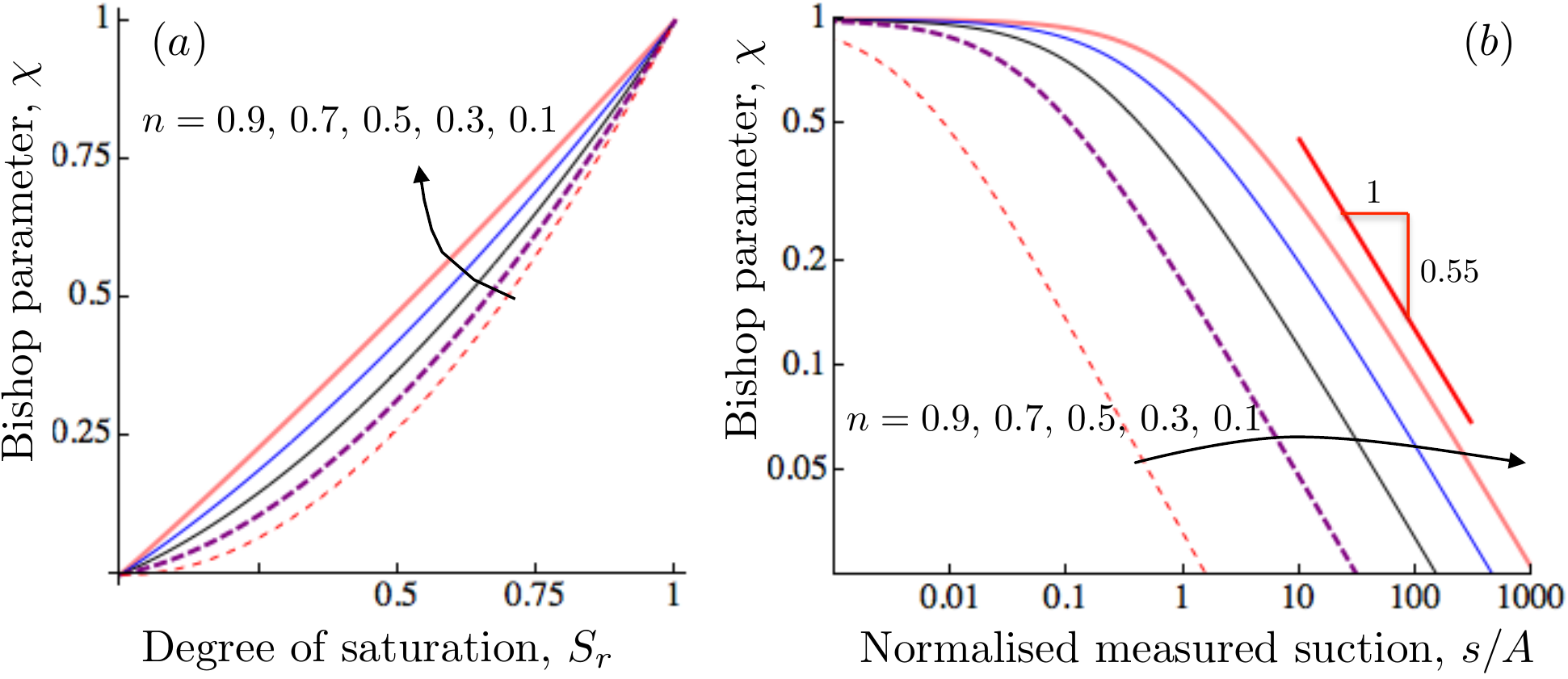}
\caption{The effect of porosity $n$ on Bishop parameter $\chi$, for fixed $\alpha=1.75$ and $\beta=2$. (a) Bishop parameter plotted in terms of the degree of saturation; (b) Bishop parameter plotted in terms of measured suction, agreeing with the relation of \cite{Khalili98} that has an empirical asymptote of 0.55.}\label{fig5}
\end{figure}

\begin{figure}[H]
\centering
\includegraphics[scale=.65]{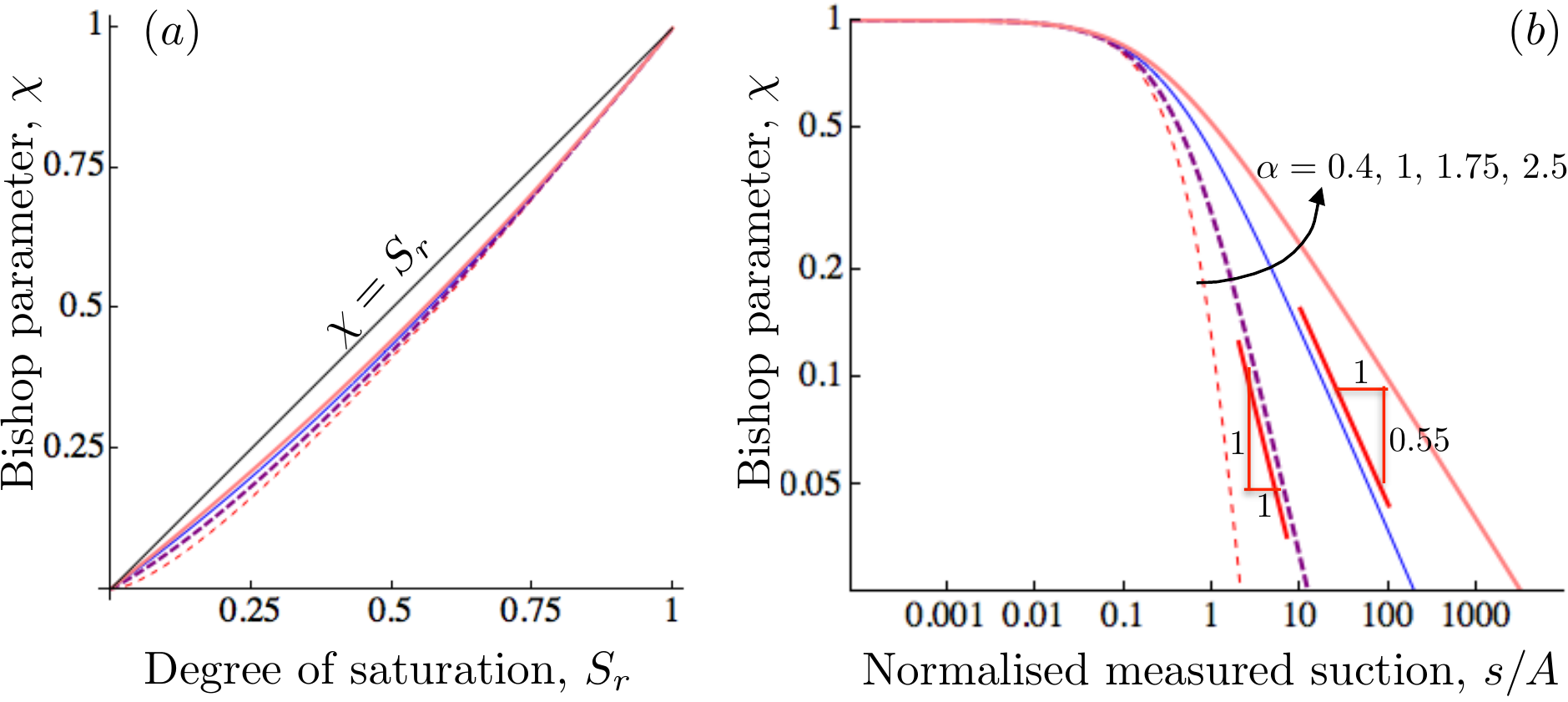}
\caption{The effect of $\alpha$ on Bishop parameter $\chi$, for fixed $\beta=2$ and $n=0.25$. (a) Bishop parameter plotted in terms of the degree of saturation; (b) Bishop parameter plotted in terms of measured suction, which can show a range of asymptotes such as the empirical slope of 0.55 in \cite{Khalili98} when $\alpha\approx 1.75$ or a slope of 1 in \cite{Aitchison85} when $\alpha\approx 1$.}\label{fig6}
\end{figure}

Recall that Figs.(\ref{fig4}) and (\ref{fig6}) were both plotted for constant $\phi_S=1-n=0.75$. However, since in this model $\chi$ depends on the porosity $n$ (and not only on either $S_r$ or $s$), it is useful to draw attention to Fig.~(\ref{fig5}). Since the soil porosity can vary during loading, the actual state cross a number of constant porosity lines in Fig.~(\ref{fig5}) during material loading. This is consistent with the recent empirical model of \cite{Alonso10} and can explain the observations of \cite{JenBurland62}.

Finally, it is useful to evaluate the difference between the measured s and intrinsic $\hat s$ suctions, since this distinction as highlighted in Eq.~(\ref{eq:twosuctions}), has been neglected from most previous work dealing with the definition of effective stress. For example, Fig.~\ref{fig7}a demonstrates their difference for the bulk modulus of water listed in Eq.~(\ref{eq:Ks}) and the  SWRC parameters $A=1\text{ MPa}$, $\alpha=1.75$, and $\beta=2$, giving a realistic measured soil-water retention curves.  As highlighted by their ratio, their values can be more than one order of magnitude apart. Previous constitutive models did not distinguish between intrinsic and measured suctions. The current paper therefore offers modellers a way to map SWRC parameters (connected to measured suction) to calculate the corresponding intrinsic suction, and thus obtaining the right effective stress. 

\begin{figure}[H]
\centering
\includegraphics[scale=.46]{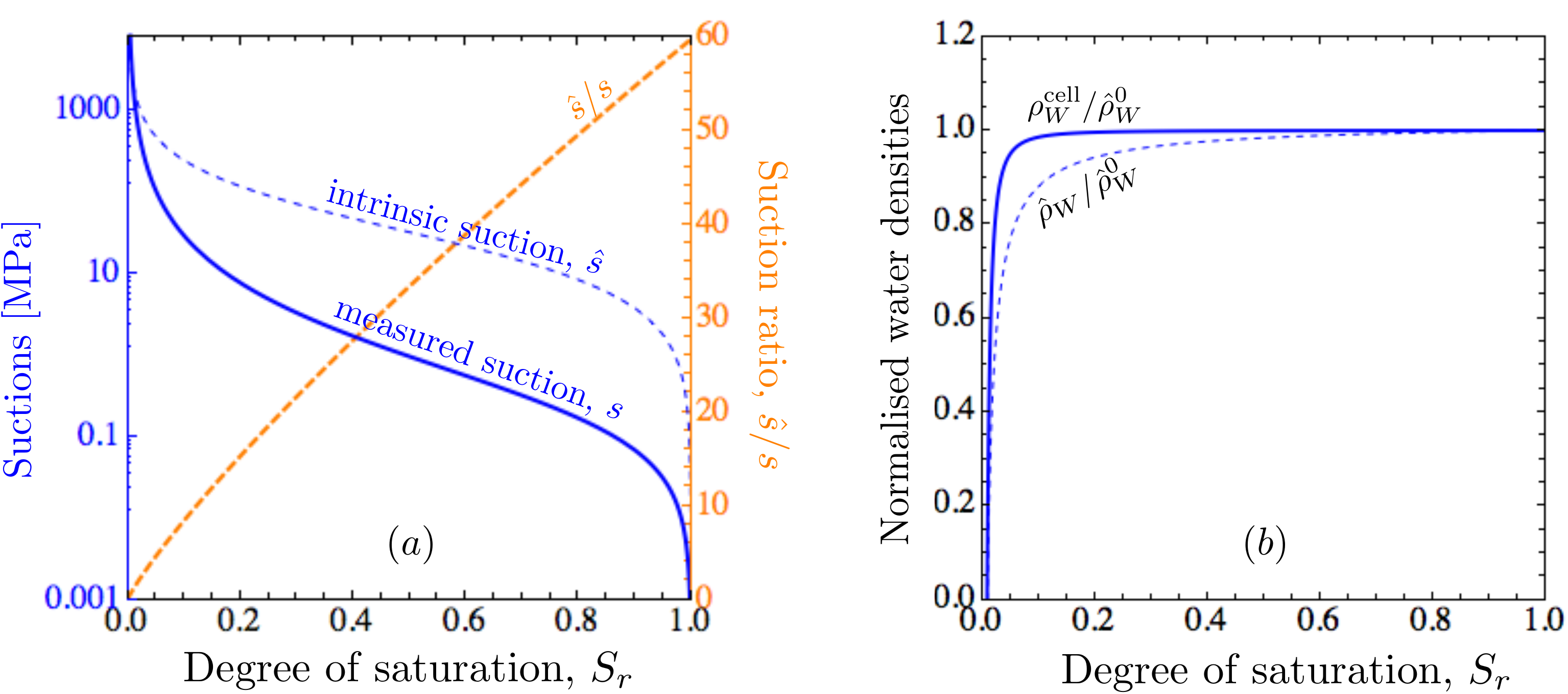}
\caption{The values of measured and intrinsic properties: (a) difference between measured s and intrinsic $\hat s$ suctions for realistic soil-water characteristics, and their ratio $\hat s/s$; (b) difference between the water density in the measurement cell $\varrho_W^\text{cell}$ and intrinsically within the soil $\hat\varrho_W$, both normalised by the  intrinsic water density when no suction applies $\hat\varrho_W^0$.}\label{fig7}
\end{figure}

For the same realistic parameters it is also possible to assess the difference between the intrinsic water density in the soil and the water density in the cell due to these suctions. Specifically, using Eqs.(\ref{33}) and (\ref{intS}), and the fact $s=-\Delta u_W$ for this case, we get $\Delta\hat\varrho_W/\hat\varrho_W^0=-\hat s/K_W$ and $\Delta\varrho_W^\text{cell}/\hat\varrho_W^0=-s/K_W$. The densities can therefore be plotted in Fig.~\ref{fig7}b as $\hat\varrho_W=\hat\varrho_W^0+\Delta\hat\varrho_W$ and $\varrho_W^\text{cell}=\hat\varrho_W^0+\Delta\hat\varrho_W^\text{cell}$. It is shown that the water in the soil expands more than in the measurement cell due to the suction. Furthermore, while the results in this figure are based on calculations using the second order approximation scheme, those were also confirmed using the exact solution. For all practical purposes, the exact and approximated solutions are identical, and thus the exact solution was not added to the figure, as it overlaps the approximated solution.

\section{Conclusions}
This paper advances the treatment of partially saturated soils through rigorous thermodynamic principles, which unravel the structure of effective stress. In soil mechanics the effective stress of soils is typically written as $\sigma_{ij}^{eff}=\sigma_{ij}-P_T \delta_{ij}$, with $\sigma_{ij}$ being the total stress and $P_T$ some sort of pressure. As proven in this paper this superposition is generally correct for any degree of saturation, if $\sigma_{ij}^{eff}$ and $P_T$ are to be interpreted as the elastic stress and thermodynamic pressure. The thermodynamic pressure can generally be written in a Bishop form $P_T=u_A+\chi(u_A-u_W)$, with the Bishop parameter $\chi$ written as a function of the three thermodynamic densities (of air, solid and water), in a way that is strictly linked to the measured suction and thus to the characteristics of soil-water retention curves (SWRC). 

It was shown that $\chi$ is independent on air density if the SWRC is also independent on air density, as frequently assumed in the literature. When the SWRC is further taken to be solely dependent on water density or the degree of saturation $S_r$, the structure of $\chi$ developed in this paper agrees with previously assumed empirical relations. However, since the characteristics of SWRC is generally known to be a function of the solid density and porosity $n$, the value of $\chi$ depends on the state of $n$, which can vary most strongly during compression. This effect of $n$ on $\chi$ is mostly neglected from most of empirical relations of $\chi$, and can therefore explain why Bishop effective stress principle was so far unable to capture both shear failure and volumetric responses. 

The derivation in this paper has adopted only a minimal number of realistic working assumptions. Without including those assumptions the thermodynamic pressure will not obey Bishop's assumption of $P_T=u_A+\chi(u_A-u_W)$, but in most practical cases Bishop seems to have got it right. For example, without using the simplifying step of neglecting air bulk modulus compared to the solid and water moduli, the thermodynamic pressure will include further terms related to compressibilities, as envisaged for fully saturated soils \cite{Lade97} with extremely soft particles. Including temperatures will further affect the structure of $P_T$. For example, in strongly sheared granular media one must consider the role of granular temperature, which can significantly elevate the value of $P_T$, as was already shown for dry media \cite{granR2,granRexp}. Other factors that we aim to study in the future include the role of cohesion in soils with very low $S_r$, the effect of mass transfer between the domains for example, through evaporation and chemical reactions, and the physics of hystereses of soil-water retention curves.

\section{Acknowledgement}
We would like to thank Giuseppe Buscarnera, Abbas El-Zein, Yixiang Gan and Adrian Russell, for fruitful discussions and helping us to gather important background in the subject field. IE acknowledges the Australian Council Research for fundings DP120104926 and DP130101291. 


\begin{appendices}
\section{Stress decomposition}
\label{sec:AppA}
The following general thermodynamic derivation for partially saturated porous soil media is consistent with the Granular Solid Hydrodynamics (GSH) framework of Jiang and Liu \cite{granR2} for dry granular materials. We consider equilibrium and take the conserved energy $w\equiv w(\mathcal{s},\varrho, c_W,c_A,g_i,\varepsilon^{e}_{ij})$ to depend on entropy $\mathcal{s}$, thermodynamic density $\varrho$, concentrations $c_A,c_W$, momentum density $g_i$,
and the elastic strain $\varepsilon^{e}_{ij}$. Using the notation of Eqs. (\ref{10}) and (\ref{StressDecom}), we write 
\begin{equation}\label{AppEnergy}
{\rm d}w=T{\rm d}\mathcal{s}+v_i{\rm d}g_i+\mu{\rm d}\varrho +\chi_A{\rm d}c_A +\chi_W{\rm d}c_W-\sigma^{eff}_{ij}{\rm d}\varepsilon^{e}_{ij},
\end{equation}
implying the definitions
\begin{align}
T\equiv\partial w/\partial \mathcal{s}, \quad v_i\equiv\partial
w/\partial g_i=g_i/\varrho,\quad \mu\equiv\partial
w/\partial\varrho
\\ 
\chi_A\equiv\partial w/\partial c_A,\quad
\chi_W\equiv\partial w/\partial c_W,\quad 
\sigma^{eff}_{ij}\equiv-\partial w/\partial \varepsilon^{e}_{ij},
\end{align}
with the pressure still given as $P_T=-w+T\mathcal{s}+\mu\varrho+v_ig_i$. We note the usual conversion $f=w-T\mathcal{s}$. Combined with Eq.(\ref{AppEnergy}), or $\nabla_iw=T\nabla_i\mathcal{s}+v_k\nabla_ig_k+\mu\nabla_i\varrho +\chi_A\nabla_ic_A +\chi_W\nabla_ic_W-\sigma^{eff}_{kj}\nabla_i\varepsilon^{e}_{kj}$,  we have
\begin{equation}\label{PressureGradient}
\nabla_iP_T=\varrho\nabla_i\mu+\mathcal{s}\nabla_iT+g_k\nabla_iv_k-\chi_A\nabla_ic_A-\chi_W\nabla_ic_W+\sigma^{eff}_{kj}\nabla_i\varepsilon^{e}_{kj}.
\end{equation}

In equilibrium, there is no dissipation and no diffusion (and no granular temperature $T_g\equiv0$). Therefore, 
the equations of motion for the energy and its variables (with $\partial_t\equiv\frac{\partial}{\partial t}$) are
\begin{align}
\label{2y-GSH} \partial_t w+\nabla_iQ_i=0, \quad
\partial_t \varrho+\nabla_i(\varrho v_i)=0,\quad \partial_t
\mathcal{s}+\nabla_i(\mathcal{s}v_i)=0,
\\\label{3y-GSH}
\partial_t g_i+\nabla_j
(\sigma_{ij}+g_iv_j)=0,
\quad (\partial_t+v_i\nabla_i)\varepsilon^{e}_{kj}=v_{kj},
\\\label{4y-GSH}
(\partial_t+v_i\nabla_i)c_A=0,\qquad (\partial_t+v_i\nabla_i)c_W=0.
\end{align}
First, we note that,  with $d_t\equiv\partial_t+v_i\nabla_i$ and $d_tc_A=d_t(\varrho_A/\varrho)=0$, we have $d_t\varrho_A/\varrho_A=d_t\varrho/\varrho=\nabla_i v_i$, or $\partial_t \varrho_A+\nabla_i(\varrho_A v_i)=0$, same as for $\varrho$. 
In other words, the continuity equation implies that the concentrations $c_A,c_W$ do not change at all, if one follows a volume element with the flux. Second, the elastic strain $\varepsilon^{e}_{ij}$ changes as usual with the deformation rate $v_{ij}\equiv\frac12(\nabla_iv_j+\nabla_jv_i)$. Third, the energy and momentum flux, $Q_i$ and $\sigma_{ij}$, need to be determined. We define $\sigma_{ij}\equiv\bar\sigma_{ij}+P_T\delta_{ij}$ with no loss of generality and determine $\bar\sigma_{ij}$.

Differentiating Eq.(\ref{AppEnergy}), we get ${\partial_ t}w = T{\partial_t} \mathcal{s} +\mu {\partial_t}\varrho + v_{i}{\partial_ t} g_{i} +\chi_A{\partial_ t}c_A +\chi_W{\partial_ t}c_W - \sigma^{eff}_{ij} {\partial_
t} \varepsilon^{e}_{ij}$, while noting that 
\begin{align*}
-\mu\partial_t\varrho&=\mu\nabla_i(\varrho v_i)=\nabla_i(\mu\varrho v_i)-\varrho v_i\nabla_i\mu,
\quad(\text{same for\,\,} -T\partial_t \mathcal{s}),
\\
-v_i\partial g_i&=v_i\nabla_j(\bar\sigma_{ij}+P_T)=\nabla_j(v_i\bar\sigma_{ij})-\bar\sigma_{ij}\nabla_jv_i+v_i\nabla_jP_T,
\end{align*}
and using Eqs. (\ref{PressureGradient},\ref{2y-GSH},\ref{3y-GSH},\ref{4y-GSH}), and $\bar\sigma_{ij}=\bar\sigma_{ji}$, we obtain
\begin{equation}
\label{xx1}
\nabla_iQ_i=\nabla_i[(T\mathcal{s}+\mu\varrho+g_kv_k) v_i
+\bar\sigma_{ij}v_j]-(\bar\sigma_{ij}-\sigma^{eff}_{ij})v_{ij}.
\end{equation}
Therefore, since $v_{ij}$ is arbitrary, 

\begin{equation}
\bar\sigma_{ij}=\sigma^{eff}_{ij},\,\,\,\,\,  Q_i=(w+P_T)v_i+\bar\sigma_{ij}v_j,
\end{equation}
thus giving 

\begin{equation}
\sigma_{ij}\equiv\sigma^{eff}_{ij}+P_T\delta_{ij},
\end{equation}
which confirms Eq.(\ref{StressDecom}).

We note that this derivation is brief and incomplete. First, convective nonlinearities and dissipative terms are not included. Second, the uniqueness of the separation as given in Eq.(\ref{xx1}) has not been shown. Third, we assume that the elastic strain and the density are independent. This is not quite true in highly compressed porous media, in which the prefactor of $P_T$ in the effective stress equation,  Eq.(\ref{StressDecom}), is no longer one even in the fully saturated case, {\it i.e.}, the Terzaghi expression Eq.(\ref{18}) does not generally hold (as conceived by various authors and summarised in \cite{Lade97}). But the essence of the derivation can be seen here, and the interested reader is referred to the literature on GSH ({\it e.g.}, see \cite{granR2}) for further details.

\section{Deriving Eqs.(23) and (24)\label{sec:AppB}}
We first note that the chemical potential is
\begin{align}
\mu_\beta&=\frac{\partial f}{\partial\rho_\beta}=\sum_\alpha\frac{\partial }{\partial\rho_\beta}\left(\hat f_\alpha\frac{\rho_\alpha}{\hat\rho_\alpha}\right)\\
&=\frac{\hat f_\beta}{\hat\rho_\beta}+\sum_\alpha\frac{\rho_\alpha}{\hat\rho_\alpha} \frac{\partial\hat f_\alpha}{\partial\hat\rho_\alpha}\frac{\partial\hat\rho_\alpha}{\partial\rho_\beta}-\sum_\alpha\frac{\rho_\alpha}{\hat\rho_\alpha}\frac{\hat f_\alpha}{\hat\rho_\alpha}\frac{\partial\hat\rho_\alpha}{\partial\rho_\beta}\\
&=\hat\mu_\beta-\frac{\hat P_\beta}{\hat\rho_\beta}+\sum_\alpha\frac{\rho_\alpha}{\hat\rho_\alpha^2}(\hat\rho_\alpha\hat\mu_\alpha-f_\alpha)\frac{\partial\hat\rho_\alpha}{\partial\rho_\beta}\\
&=\hat\mu_\beta-\frac{\hat P_\beta}{\hat\rho_\beta}+\sum_\alpha\frac{\rho_\alpha}{\hat\rho_\alpha^2}\hat P_\alpha\frac{\partial\hat\rho_\alpha}{\partial\rho_\beta}.
\end{align}
Turning next to the pressure, we find
\begin{align}
P_T=\rho\frac{\partial f}{\partial\rho}-f=\sum_\beta\rho_\beta\frac{\partial f}{\partial\rho_\beta}-f=\sum_\beta\rho_\beta\mu_\beta-\frac{\rho_\beta}{\hat\rho_\beta}\hat f_\beta.
\end{align}
Inserting the expressions for $\mu_\beta$, Eqs.(85,87), we have 
\begin{align}
P_T=\sum_{\alpha,\beta}\frac{\rho_\beta\rho_\alpha}{\hat\rho_\beta^2}{\hat P_\beta}\frac{\partial\hat\rho_\beta}{\partial\rho_\alpha}=\sum_{\beta} \frac{\rho_\beta}{\hat\rho_\beta^2}{\hat P_\beta}\sum_\alpha\rho_\alpha\frac{\partial\hat\rho_\beta}{\partial\rho_\alpha}
=\sum_{\beta} \frac{\rho_\beta}{\hat\rho_\beta^2}{\hat P_\beta}\left(\rho\frac{\partial\hat\rho_\beta}{\partial\rho}\right).
\end{align}
Finally, employing Eqs.(25), we arrive at the respective second equality sign of every line of Eqs.(23) and (24).
\end{appendices}

\end{document}